\documentclass[trackchanges, twocolumn]{aastex701}

\newcommand{\cahk}{Ca \scriptsize{\uppercase\expandafter{\romannumeral2}} \normalsize H$\&$K }
\usepackage{amsmath}
\usepackage{chngcntr}
\usepackage[figuresright]{rotating}
\usepackage{ctable}
\usepackage{float}
\usepackage{adjustbox}
\usepackage{graphicx}
\usepackage{realboxes}
\usepackage{epsfig}
\usepackage{threeparttablex, tablefootnote}
\usepackage{subfigure}
\usepackage{graphicx}
\usepackage{CJK}
\usepackage{booktabs}
\usepackage{multirow}
\usepackage{amsmath}
\usepackage{url}
\usepackage{hyperref}  

\providecommand{\dodoi}[1]{%
  \href{https://doi.org/#1}{doi:\nolinkurl{#1}}%
}

\begin{document}


\title{Varying core-envelope coupling efficiency identified from stellar rotation--activity relation}

\author[orcid=0000-0003-3474-5118,gname=Henggeng, sname='Han']{Henggeng Han} 
\affiliation{National Astronomical Observatories, Chinese Academy of Sciences, Beijing 100101, People's Republic of China}
\email{hghan@nao.cas.cn}

\author[orcid=0000-0003-3116-5038,gname=Song, sname='Wang']{Song Wang} 
\altaffiliation{Corresponding author: songw@bao.ac.cn}
\affiliation{National Astronomical Observatories, Chinese Academy of Sciences, Beijing 100101, People's Republic of China}
\affiliation{Institute for Frontiers in Astronomy and Astrophysics, Beijing Normal University, Beijing, 102206, People's Republic of China}
\email{songw@bao.ac.cn}

\author[orcid=0000-0003-4917-7221,gname=Huiqin, sname='Yang']{Huiqin Yang} 
\affiliation{National Astronomical Observatories, Chinese Academy of Sciences, Beijing 100101, People's Republic of China}
\affiliation{Institute for Frontiers in Astronomy and Astrophysics, Beijing Normal University, Beijing, 102206, People's Republic of China}
\email{yhq@nao.cas.cn}

\author[orcid=0009-0007-4501-4376,gname=Xue, sname='Li']{Xue Li} 
\affiliation{School of Astronomy and Space Science, University of Chinese Academy of Sciences, Beijing 100049, People's Republic of China}
\email{lixue@bao.ac.cn}

\author[orcid=0009-0006-7556-8401,gname=Chuanjie, sname='Zheng']{Chuanjie Zheng} 
\affiliation{School of Astronomy and Space Science, University of Chinese Academy of Sciences, Beijing 100049, People's Republic of China}
\email{zhengchuanjie@ucas.ac.cn}

\author[gname=Xiangyu, sname='Li']{Xiangyu Li} 
\affiliation{National Astronomical Observatories, Chinese Academy of Sciences, Beijing 100101, People's Republic of China}
\affiliation{School of Astronomy and Space Science, University of Chinese Academy of Sciences, Beijing 100049, People's Republic of China}
\email{lixy@bao.ac.cn}

\author[orcid=0000-0001-5941-3246,gname=Cunshi, sname='Wang']{Cunshi Wang} 
\affiliation{School of Astronomy and Space Science, University of Chinese Academy of Sciences, Beijing 100049, People's Republic of China}
\email{wangcunshi@nao.cas.cn}

\author[orcid=0000-0002-2874-2706,gname=Jifeng, sname='Liu']{Jifeng Liu} 
\affiliation{National Astronomical Observatories, Chinese Academy of Sciences, Beijing 100101, People's Republic of China}
\affiliation{School of Astronomy and Space Science, University of Chinese Academy of Sciences, Beijing 100049, People's Republic of China}
\affiliation{Institute for Frontiers in Astronomy and Astrophysics, Beijing Normal University, Beijing, 102206, People's Republic of China}
\affiliation{New Cornerstone Science Laboratory, National Astronomical Observatories, Chinese Academy of Sciences, Beijing, 100012, People's Republic of China}
\email{jfliu@bao.ac.cn}


\begin{abstract}

Core-envelope coupling provides a reasonable explanation of the spin-down stalling of stars in open clusters, which was not predicted by classical gyrochronology. However, it remains an open question whether the coupling efficiency is constant or variable. M dwarfs, possessing thicker convective envelopes and thus longer coupling timescales than other late-type stars, are ideal objects for this investigation.
In this work, based on the $R_{\rm{HK}}^{'}$ measurements from LAMOST and DESI spectra, we construct new rotation--activity relations for M dwarfs. Unlike the traditional picture, we suggest that the new relation consists of three distinct regimes of fast, intermediate, and slow rotation, closely matching the three sequences of gyrochronology, namely the ``Convective'' sequence, ``Gap'', and ``Interface'' sequence. Our study reveals, for the first time, a variable activity decay rate in the intermediate-rotation regime (i.e., the ``Gap'' region). This implies a varying core-envelope coupling efficiency, peaking towards the end of this region. It also coincides with the well-known stage of stalled stellar spin-down.

\end{abstract}



\section{Introduction}
\label{sec:intro}

Late-type stars experience gradual angular momentum loss through magnetized stellar winds \citep[e.g.][]{1988ApJ...333..236K}. This suggests stellar rotation can trace their age, which was referred as gyrochronology \citep{2003ApJ...586..464B, 2016ApJ...823...16B}. The foundation of modern gyrochronology was established by \cite{2003ApJ...586..464B}, who connected rotational evolution to magnetic dynamo through three sequences: the ``Convective" sequence (``C'' sequence), the ``Interface" sequence (``I'' sequence) and the transition region, i.e., the ``Gap''. Subsequently, gyrochronology has been systematically calibrated using stellar clusters of various ages \citep[e.g.,][]{2009ApJ...695..679M, 2011ApJ...733..115M, 2015Natur.517..589M}.

\begin{figure*}
\centering
\subfigure[]{
\includegraphics[width=0.85\textwidth]{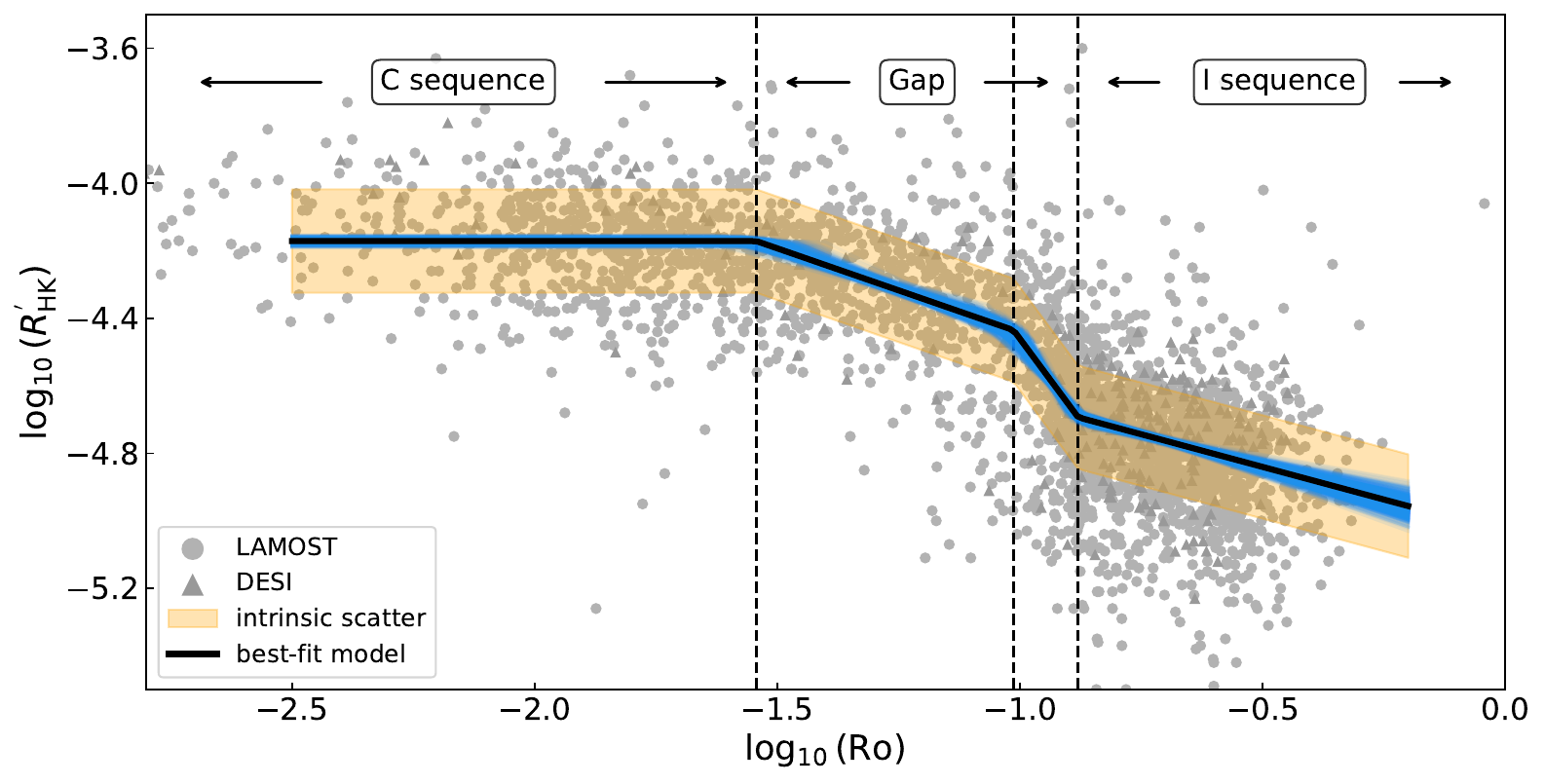}}
\subfigure[]{
\includegraphics[width=0.85\textwidth]{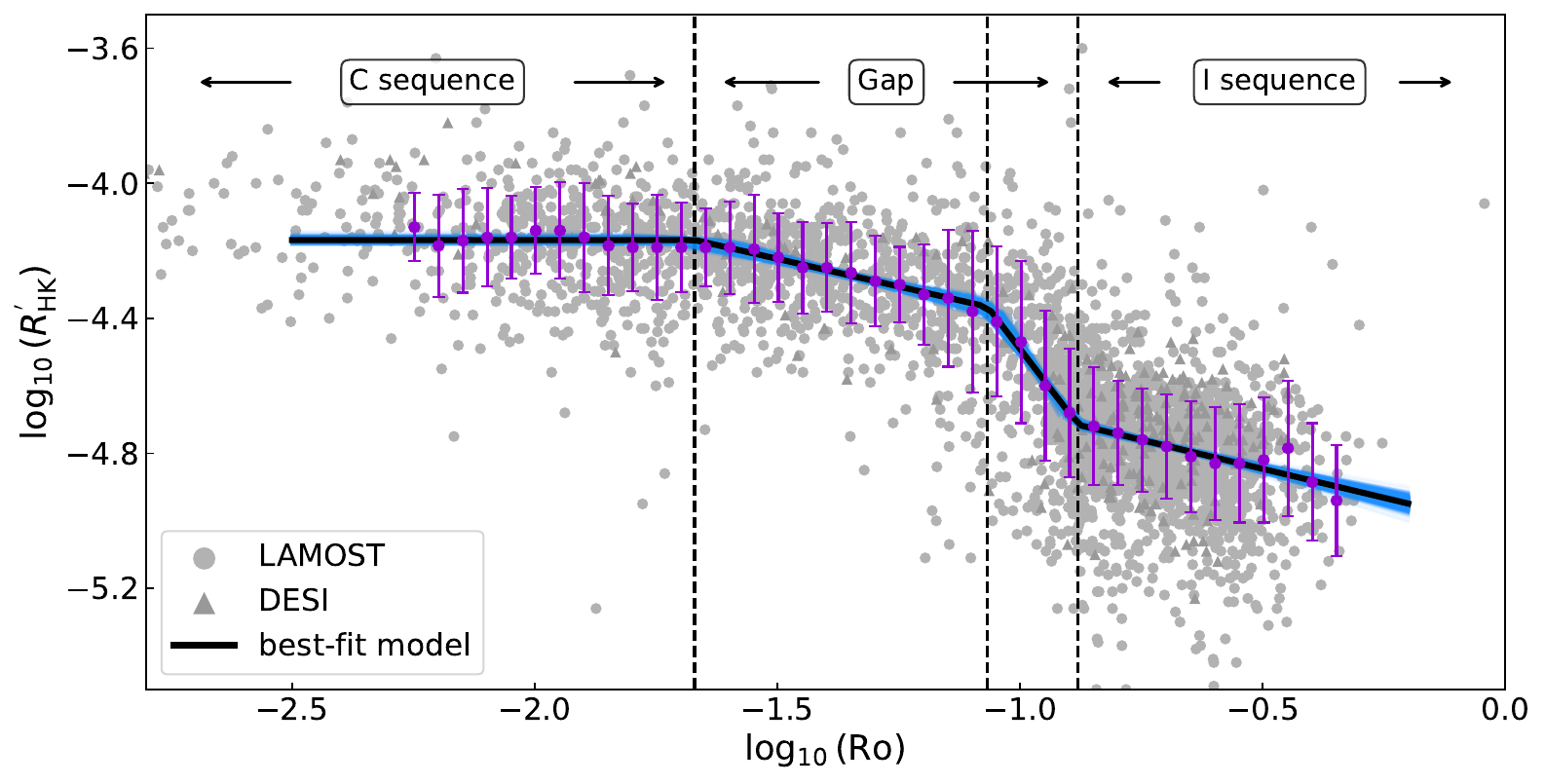}}
\caption{Panel (a): $R_{\rm{HK}}^{'}$--Ro relation of M dwarfs. Black line is the best-fit model. Black vertical dashed lines mark the three knee points. Blue region is composed by 1000 models randomly extracted from posterior probability distributions. Orange area represents intrinsic scatter of the log$_{10}(R_{\rm{HK}}^{'})$. Panel (b): Same as panel (a) but for binned median fitting method.}
\label{ar_fit_piece.fig}
\end{figure*}

However, currently the gyrochronology is not valid for all evolutionary phases. For example, studies of old field stars show their rotation periods deviate from predictions, supporting the idea of weakened magnetic braking scenario \citep{2016Natur.529..181V}. Data from open clusters suggest that stars will settle onto a well-defined, slow-rotating sequence and the time it takes to reach this sequence depends on the stellar masses, which is not predicted by gyrochronology. By an age of about 1 Gyr, stars more massive than $\approx$ 0.6\,$M_{\rm{\odot}}$ have all converged to form a distinct, slow-rotating sequence (Figure 1 of \cite{2003ApJ...586..464B} and \cite{2020A&A...636A..76S}). Meanwhile, \cite{2020ApJ...904..140C} proposed that stars with $M < 0.55\, M_{\rm{\odot}}$ may stall for at least 1.3 Gyr, which has been anticipated by \cite{2011ApJ...733L...9M} using \emph{Kepler} data of NGC 6811.

Such phenomenon can be explained by the core-envelope coupling process, which will redistribute angular momentum in the stellar interior. Early works that incorporated this mechanism established rotational models for solar-type stars \citep[e.g.][]{1991ApJ...376..204M, 1997ApJ...480..303K}; these models were later extended and refined for lower-mass stars \citep[e.g.][]{2010ApJ...716.1269D, 2013A&A...556A..36G, 2015A&A...577A..98G}. Recently, \cite{2020A&A...636A..76S} developed a two-zone rotational evolution model with a mass-dependent coupling timescale to explain the observed spin-down stalling. Some studies \cite[e.g.,][]{2022AJ....164..251L} also proposed that the period gap seen in the color-period diagram of \emph{Kepler} targets was due to the core-envelope coupling process.

The evolution of stellar rotation will further lead to changes in stellar magnetic activities, and the rotation--activity relation has also been divided into different regions to link to the gyrochronology. \cite{2003ApJ...586L.145B} first linked the ``C'' sequence, the ``Gap'', and the ``I'' sequence to the supersaturation, saturation, and decay regions of rotation--activity relations, respectively. 
The supersaturation regime, where magnetic activity declines with increasing rotation for extremely fast‑rotating stars, was first identified decades ago \citep{1996A&A...313..815R}, yet its physical reality remains debated \citep[e.g.,][]{2000MNRAS.318.1217J,2011MNRAS.411.2099J,2024ApJS..273....8H,2025MNRAS.542.2431L}.
\cite{2011ApJ...743...48W} found that stars in the supersaturation region were far fewer than those in the saturation region, and then associated the ``C'' sequence with the saturation region. This interpretation, however, introduced a discrepancy: the transition between the saturation and unsaturated regimes in the rotation--activity relation appears smooth, contrary with the distinct ``Gap'' phase predicted by gyrochronology. Recently, \cite{2025A&A...699A.251Y} constructed the rotation--activity relation using a linear--log scaling, successfully identifying the ``Gap'' region.


\begin{table*}
\begin{center}
{\scriptsize
  \setlength{\tabcolsep}{1pt}
\caption{Stellar parameters, stellar activity indices, rotation periods, and convective turnover times for objects with LAMOST observations.}
\label{tab:table1}
\begin{tabular}{lcccccccccccc}
\toprule
ID & R.A. & Decl & $T_{\rm{eff}}$ & log$g$ & [Fe/H] & $S_{\rm{MW}}$ & $(B-V)_{0}$ & $(BP-RP)_{0}$ & Distance & $\text{log}_{10}(R_{\text{HK}}^{'})$ & $P_{\rm{rot}}$ & $\tau_{c}$ \\
 & degree & degree & (K) & (dex) & (dex) & & (mag) & (mag) & (pc) & & (day) & (day) \\
 (1) & (2) & (3) & (4) & (5) & (6) & (7) & (8) & (9) & (10) & (11) & (12) & (13) \\ 
\hline
\midrule
3804658125186560 & 46.27821 & 4.89415 & 3673$\pm$43 & 4.85$\pm$0.05 & 0.24$\pm$0.09 & 10.18$\pm$2.73 & 1.48 & 2.13 & 383$\pm$22 & $-$4.04$\pm$0.12 & 8.25 & 102.40$\pm$83.82 \\
6087931459094400 & 41.16184 & 5.17798 & 3977$\pm$37 & 4.65$\pm$0.04 & $-$0.05$\pm$0.08 & 4.40$\pm$0.80 & 1.38 & 1.68 & 686$\pm$51 & $-$4.25$\pm$0.08 & 3.30 & 103.85$\pm$11.41 \\
8772492177548160 & 44.95876 & 8.36457 & 3424$\pm$66 & 4.71$\pm$0.09 & 0.09$\pm$0.14 & 1.35$\pm$0.49 & 1.54 & 2.28 & 185$\pm$3 & $-$4.98$\pm$0.16 & 49.31 & 176.39$\pm$25.65 \\
8777336900638080 & 44.85226 & 8.47707 & 3522$\pm$17 & 4.75$\pm$0.03 & $-$0.05$\pm$0.04 & 1.13$\pm$0.12 & 1.51 & 2.30 & 95$\pm$1 & $-$5.02$\pm$0.05 & 53.81 & 177.12$\pm$31.09 \\
... & ... & ... & ... & ... & ... & ... & ... & ... & ... & ... & ... & ... \\

\bottomrule
\end{tabular}}
\end{center}
\end{table*}

\begin{table*}
\begin{center}
{\scriptsize
  \setlength{\tabcolsep}{0.9pt}
\caption{Stellar parameters, stellar activity indices, rotation periods, and convective turnover times for objects with DESI observations.}
\label{tab:table2}
\begin{tabular}{lcccccccccccc}
\toprule
ID & R.A. & Decl & $T_{\rm{eff}}$ & log$g$ & [Fe/H] & $S_{\rm{MW}}$ & $(B-V)_{0}$ & $(BP-RP)_{0}$ & Distance & $\text{log}_{10}(R_{\text{HK}}^{'})$ & $P_{\rm{rot}}$ & $\tau_{c}$ \\
 & degree & degree & (K) & (dex) & (dex) & & (mag) & (mag) & (pc) & & (day) & (day) \\
 (1) & (2) & (3) & (4) & (5) & (6) & (7) & (8) & (9) & (10) & (11) & (12) & (13) \\ 
\hline
\midrule
129012884770564608 & 43.29754 & 29.37426 & 3666$\pm$3 & 4.69$\pm$0.01 & $-$0.11$\pm$0.01 & 7.00$\pm$0.42 & 1.49 & 2.18 & 95$\pm$1 & $-$4.20$\pm$0.03 & 4.58 & 132.87$\pm$25.95 \\
604706862424426496 & 133.25534 & 11.52213 & 3919$\pm$3 & 4.86$\pm$0.01 & $-$0.19$\pm$0.01 & 1.92$\pm$0.19 & 1.40 & 1.92 & 1054$\pm$313 & $-$4.65$\pm$0.04 & 35.77 & 119.27$\pm$29.56 \\
609556331473610880 & 130.49931 & 14.73676 & 3305$\pm$5 & 4.92$\pm$0.02 & $-$0.02$\pm$0.02 & 9.32$\pm$0.66 & 1.60 & 2.63 & 211$\pm$7 & $-$4.21$\pm$0.03 & 3.43 & 493.03$\pm$66.64 \\
656258981217960448 & 125.14409 & 16.98324 & 3465$\pm$8 & 4.51$\pm$0.04 & $-$0.27$\pm$0.03 & 1.27$\pm$0.47 & 1.52 & 2.36 & 289$\pm$12 & $-$5.00$\pm$0.16 & 30.99 & 156.89$\pm$8.02 \\
... & ... & ... & ... & ... & ... & ... & ... & ... & ... & ... & ... & ... \\

\bottomrule
\end{tabular}}
\end{center}
\end{table*}

The core-envelope coupling process is thought to occur in the ``Gap'' region \citep{2025A&A...699A.251Y}. However, whether the efficiency of the coupling process is constant or not is still an open question. M dwarfs exhibit much thicker convective envelopes than other late-type stars. As a result, they experience longer timescales in the ``Gap'' phase \citep{2025A&A...699A.251Y}, making them ideal targets to investigate the core-envelope coupling process. 
In this work, we use spectra from the Large Sky Area Multi-Object Fiber Spectroscopic Telescope (LAMOST; \citealt{2012RAA....12.1197C, 2015RAA....15.1095L}) and Dark Energy Spectroscopic Instrument (DESI; \citealt{desicollaboration2025datarelease1dark}) to study the rotation--activity behaviour in the ``Gap'' region. 
We successfully identify a region with a sharp decrease of activity levels, possibly suggesting an increasing efficiency of core-envelope coupling. 
The paper is organized as follows. In Section \ref{sec:sample} we introduce the sample and data reduction. The new rotation--activity relation is presented in Section \ref{sec:relation}. In Section \ref{sec:gyro}, we establish a connection between the rotation--activity relation and gyrochronology. In Section \ref{sec:sum}, we give a brief summary.

\section{Sample and Methods}
\label{sec:sample}
\subsection{Sample cleaning}
\label{sec:data}
LAMOST is a Schmidt telescope with a field of view of 5 degrees. Its innovative focal plane design gathers 4,000 optical fibers, enabling simultaneous observation of thousands of stars in a single exposure \citep{2012RAA....12.1197C}. The instrument provides two spectral modes: low-resolution (R$\sim$1800) and medium-resolution (R$\sim$7500) spectroscopy \citep{2012RAA....12.1197C, 2020arXiv200507210L}. For this study, we employed the data release 12 (DR12) low-resolution spectra, which contains roughly 684,522 M dwarfs. 

The Dark Energy Spectroscopic Instrument (DESI), installed on the 4-meter Mayall Telescope at Kitt Peak National Observatory, is a spectrograph capable of simultaneously gathering 5,000 spectra across a resolution range of 2,000$-$5,000 in a single exposure \citep{2016arXiv161100036D}. Its primary five-year survey has mapped roughly 14,000 square degrees of sky to probe dark matter distribution, providing an ideal sample to investigate magnetic activity of halo stars, which contains roughly 722,656 M dwarfs in its DESI data release 1 (DR1) \citep{desicollaboration2025datarelease1dark}. 

Stellar parameters including effective temperature ($T_{\rm{eff}}$), surface gravity (log$g$) and metallicity ([Fe/H]) were extracted from LAMOST DR12 and DESI DR1 parameter catalogs. To pick out M dwarfs, we applied consistent selection criteria to both datasets: (1) $T_{\rm{eff}}$ $\leq$ 4,000 K and (2) log$g$ $\geq$ 3.5. Additionally, we applied signal-to-noise ratio (S/N) cuts: LAMOST spectra required $g$-band S/N larger than 10, while DESI spectra required blue-arm S/N larger than 10. For stars with multiple observations, we used the parameters derived from the highest-S/N spectrum. 

Furthermore, we applied an additional cut based on the color $(BP-RP)_{0}$.
We cross-matched our sample with the {\it Gaia} eDR3 catalog \citep{2021A&A...649A...1G} and the distance catalog \citep{2021AJ....161..147B} to derive the {\it Gaia} magnitudes and distances.
The sources with distances larger than 2 kpc or a relative parallax error exceeding 0.2 were excluded. 
We obtained interstellar reddening values from \cite{2019ApJ...887...93G}, supplemented by those from \cite{1998ApJ...500..525S}. Extinction coefficients were taken from \cite{1999PASP..111...63F}. 
We excluded targets with intrinsic color $(BP-RP)_{0} < 1.5$ to remove potentially misclassified M dwarfs \citep{2019ApJS..244....8Z,2025ApJS..277...47Z}.

Finally, to derive a clean sample of M dwarfs, We followed the cleaning processes of \cite{2024ApJ...966...69L} (see Section 2.2 of the paper for details) to remove potential contaminants, which mainly come from non-stellar objects, binaries, evolved stars and young stellar objects. We also cross-matched the sample with SIMBAD to remove sources marked as ``GALAXY'' or ``QSO''. These selection steps yielded samples of 139,087 LAMOST targets and 99,010 DESI targets.

\subsection{Activity Indices}
\label{sec:index}
\subsubsection{$S$-index}
The $S$-index, first introduced by \cite{1978PASP...90..267V}, provides a standardized measurement of stellar magnetic activity through \cahk emission. This quantity was originally derived from observations obtained with the four-channel spectrophotometer installed on the Mount Wilson Observatory 60-inch telescope. The index is formally defined as:
\begin{equation}
    S_{\rm{MW}} = \alpha \frac{N_{H} + N_{K}}{N_{V} + N_{R}}.
\end{equation}
Here $N_{H}$ and $N_{K}$ are background corrected counts of \emph{H} and \emph{K} bands centered at 3968.47 \AA \, and 3933.664 \AA, respectively. The bandpass has a triangle shape with full width at half maximum of 1.09 \AA. \, $N_{V}$ and $N_{R}$ are background corrected counts of \emph{R} and \emph{V} bands centered at 4001.067 \AA \, and 3901.067 \AA, respectively. The bandpass is a rectangle bandpass with 20 \AA \, width. $\alpha$ is a normalizing factor that was used for calibration of the $S$-indices and \cahk flux corresponding to different instruments.

\begin{figure*}
\subfigure[]{
\includegraphics[width=0.45\textwidth]{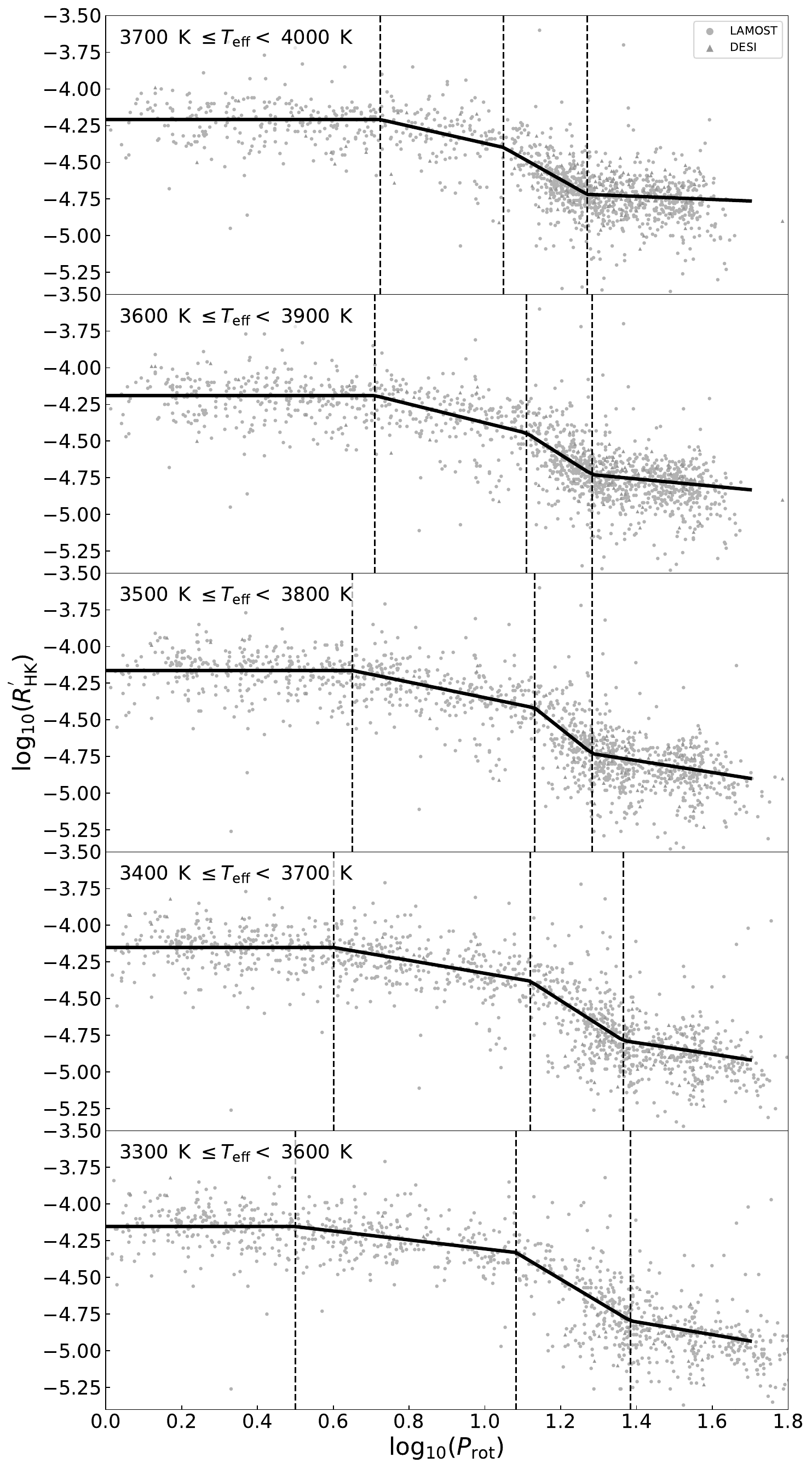}}
\subfigure[]{\includegraphics[width=0.45\textwidth]{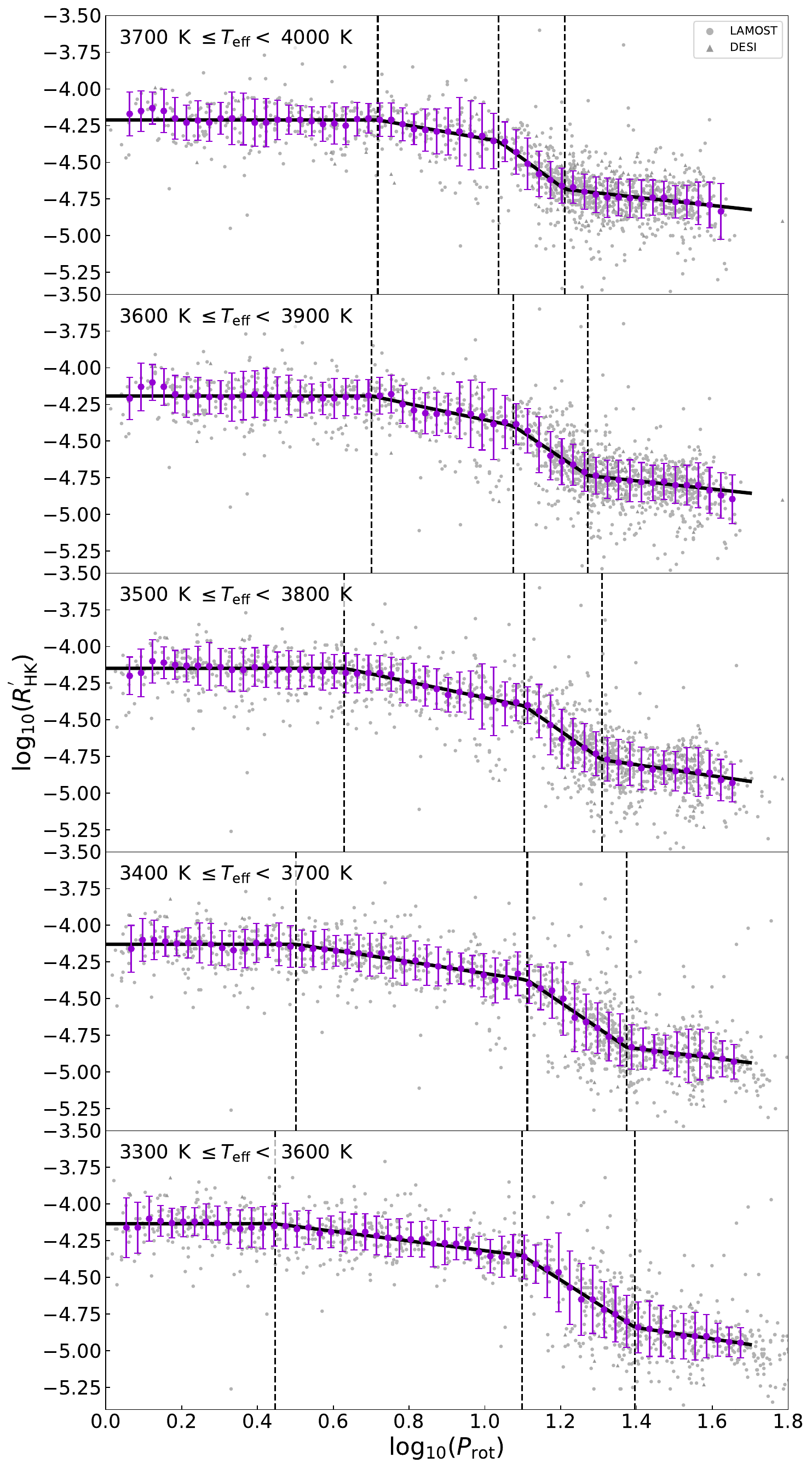}}
\caption{Panel (a): $R_{\rm{HK}}^{'}$--$P_{\rm{rot}}$ relations corresponding to various $T_{\rm{eff}}$ bins. Black lines are the best-fit models. Dashed vertical lines mark the knee points. Panel (b): $R_{\rm{HK}}^{'}-P_{\rm{rot}}$ relations using the binned median fitting method.}
\label{teff_bin.fig}
\end{figure*}

In this study, we adopted a similar method to compute the $S$-indices with $S = (H + K)/(R + V)$ and then calibrated to Mount Wilson $S$-indices ($S_{\rm{MW}}$).
Considering that the LAMOST and DESI spectra are both given in vacuum wavelength, the \emph{H} and \emph{K} are integrated flux in the triangle bandpass with 1.09 \AA \, centered at 3969.59 \AA \, and 3934.78 \AA \,, respectively. The \emph{R} and \emph{V} are the integrated flux centered at 4001 \AA \, and 3901 \AA \, respectively. In this work, no normalization factor was applied since both the LAMOST $S$-indices ($S_{\rm{LAMOST}}$) and DESI $S$-indices ($S_{\rm{DESI}}$) would be finally converted into $S_{\rm{MW}}$. We derived the $S$-index uncertainties through error propagation of the flux measurements at each wavelength.

Recently, \cite{2025ApJ...984....2H} established calibration relations between the $S$-indices from modern spectroscopic surveys (with different resolutions) and $S_{\rm{MW}}$. These calibrations were derived using PHOENIX synthetic spectra with simulated chromospheric activity, combined with the calibration between HARPS S-indices and $S_{\rm{MW}}$ from \cite{2018A&A...616A.108B}. In this work, we applied the coefficients from Table 1 of \cite{2025ApJ...984....2H} to convert both $S_{\rm{LAMOST}}$ and $S_{\rm{DESI}}$ indices to the $S_{\rm{MW}}$. We note that these calibrations are only valid for targets with [Fe/H] $> -1$, thus excluded targets with [Fe/H] $\leq -1$. Given the spectral resolutions (1800 for LAMOST and 2000 for DESI's blue arm), the conversions to $S_{\rm{MW}}$ for LAMOST and DESI targets follow
\begin{equation}
    S_{\rm{MW}} = 49.218 \times S_{\rm{LAMOST}} - 0.415, 
\end{equation}
and 
\begin{equation}
    S_{\rm{MW}} = 44.981 \times S_{\rm{DESI}} - 0.353,
\end{equation}
respectively.
Figure \ref{scomp.fig} compares the $S_{\rm{MW}}$ values derived from LAMOST and DESI for common targets in both surveys. The median difference of approximately 0.01 indicates excellent agreement between the two calibrations.

\subsubsection{Chromospheric activity index: $R_{\rm{HK}}^{'}$}
\label{sec:rhk}
Since the $S_{\rm{MW}}$ includes photospheric contribution, \cite{1984ApJ...279..763N} introduced the $R_{\rm{HK}}^{'}$ index to quantify pure chromospheric activity. The conversion was built based on the early work by \cite{1982A&A...107...31M}, who developed the color-dependent conversion factor $C_{\rm{cf}}$ to transform $S_{\rm{MW}}$ into $R_{\rm{HK}}$, thereby removing photospheric contribution from the continuum:
\begin{equation}
    R_{\rm{HK}} = 1.34 \times 10^{-4} \times C_{\rm{cf}} \times S_{\rm{MW}},
\end{equation}
where the $C_{\rm{cf}}$ was written as:
\begin{equation}
\small
    \text{log}_{10}(C_{\rm{cf}}) = 1.13(B-V)_{0}^{3} - 3.91(B-V)_{0}^{2} + 2.84(B-V)_{0} - 0.47.
\end{equation}
Then \cite{1984ApJ...279..763N} slightly modified the factor: log$_{10}(C_{\rm{cf}}^{'}) = \text{log}_{10}(C_{\rm{cf}}) + \Delta \rm{log}_{10}(\emph{C})$, where
\begin{equation}
\Delta \text{log}_{10} (C) = 
\begin{cases}
0, & \text{for } (B-V)_{0} > 0.63 \\
0.135x - 0.814x^2 \\ + 6.03x^3, & \text{for } (B-V)_{0} \leq 0.63
\end{cases}
\end{equation}
Here $x = 0.63 - (B-V)_{0}$. Furthermore, \cite{1984ApJ...279..763N} used some inactive stars to established an empirical color-dependent photospheric contribution factor, which was further subtracted from the $R_{\rm{HK}}$:
\begin{equation}
    \text{log}_{10}(R_{\rm{phot}}) = -4.898 + 1.918(B-V)_{0}^{2} - 2.893(B-V)_{0}^{3}.
\end{equation}
Then for stars with $0.4 \leq (B-V)_{0} \leq 1.6$ the pure chromospheric activity index $R_{\rm{HK}}^{'}$ was defined as:
\begin{equation}
    R_{\rm{HK}}^{'} = 1.34 \times 10^{-4} \times C_{\rm cf}^{'} \times S_{\rm{MW}} - R_{\rm{phot}}.
    \label{rhk.eq}
\end{equation}
In this study, we computed chromospheric activity indices using Equation (\ref{rhk.eq}). Stellar intrinsic colors $(B-V)_{0}$ were determined by interpolating $T_{\rm{eff}}$ with the empirical relations from \cite{2013ApJS..208....9P}. The uncertainties in $R_{\rm{HK}}^{'}$ were derived through random sampling of stellar parameters and $S$-indices based on their errors. Figure \ref{scomp.fig} shows the consistency between $R_{\rm{HK}}^{'}$ from LAMOST and DESI for common targets.

\subsection{Rotation Periods and Rossby Number}
\label{sec:prot}
Both rotation and convection play key roles in generating stellar magnetic fields. The periodic modulation of light curves due to co-rotating active regions provides a way to measure stellar rotation periods ($P_{\rm rot}$) \citep{2013MNRAS.432.1203M, 2014ApJS..211...24M, 2019ApJS..244...21S}. To build a large sample of stars with measured rotation periods, we compiled results from multiple catalogs \citep{2014ApJS..211...24M, 2019ApJS..244...21S, 2020A&A...635A..43R, 2022AJ....164..251L, 2025ApJS..276...57G}, spanning observations from the \emph{Kepler}, \emph{K}2, TESS, and ZTF surveys. For the TESS sample, we retained only stars with a classification probability greater than 0.5 to ensure reliable period measurements \citep{2025ApJS..276...57G}. In cases where rotating variables were identified in multiple catalogs, we prioritized periods based on the photometric data quality (i.e., \emph{Kepler} $>$ \emph{K}2 $>$ TESS $>$ ZTF). Furthermore, light curves were visually checked. \emph{Kepler} and \emph{K}2 light curves used in this work can be found at MAST \citep{https://doi.org/10.17909/t9488n, https://doi.org/10.17909/t9ws3r}. TESS light curves used in this work can be found at MAST \citep{https://doi.org/10.17909/t9-nmc8-f686}. ZTF light curves can be found at IPAC \citep{https://doi.org/10.26131/irsa598}.

\begin{figure}
\centering

\includegraphics[width=0.45\textwidth]{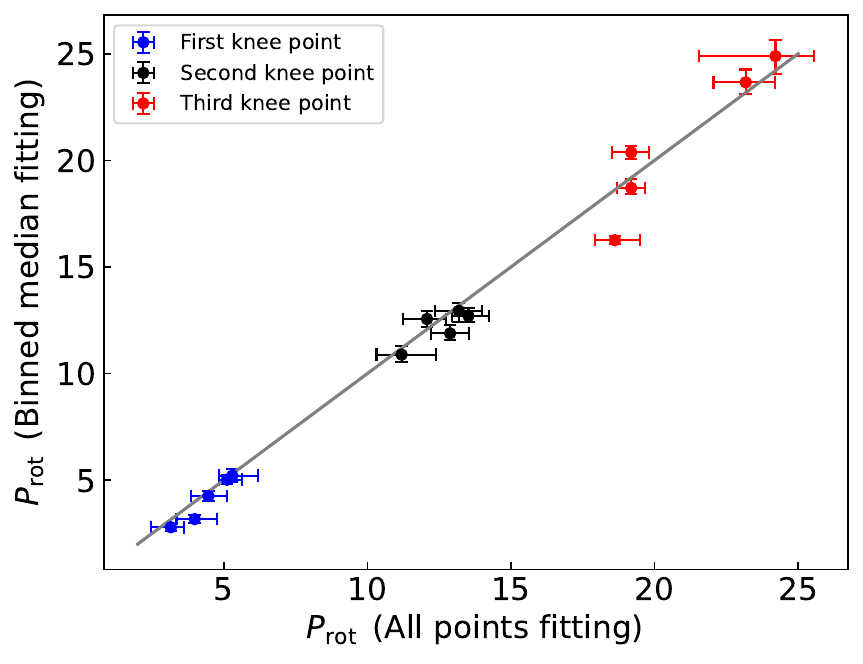}

\caption{Comparisons between the knee points in $R_{\rm{HK}}^{'}-P_{\rm{rot}}$ relations, from the fit using all data points and the fit to the binned medians.}
\label{comparison.fig}
\end{figure}

The Rossby number (Ro), defined as the ratio of the rotation period to the convection turnover time ($\tau_{c}$), is a parameter in dynamo theory that quantify the effects of rotation on convective cells \citep{1955ApJ...122..293P}. Investigating correlations between stellar activity and Ro indices can thus provide insights into dynamo mechanisms. In this work, we utilized the grids of the Yale-Potsdam Stellar Isochrones \citep{2017ApJ...838..161S} to calculate global $\tau_{\rm{c}}$. For each target, we matched model grids based on its $T_{\rm{eff}}$ and log$g$ within their typical errors (i.e., 200 K and 0.15 dex) and used the $\tau$ corresponding to the nearest model grid as the final $\tau_{c}$ and half of the 16th–84th percentile range of $\tau$ of the model grids as error of $\tau_{c}$. This process was repeated across all metallicity grids, and the final $\tau_{c}$ was determined by interpolating the [Fe/H] onto these grids. Median value of error of $\tau_{c}$ among different metallicity grids was used as the final error.

\begin{figure*}
\centering

\includegraphics[width=0.9\textwidth]{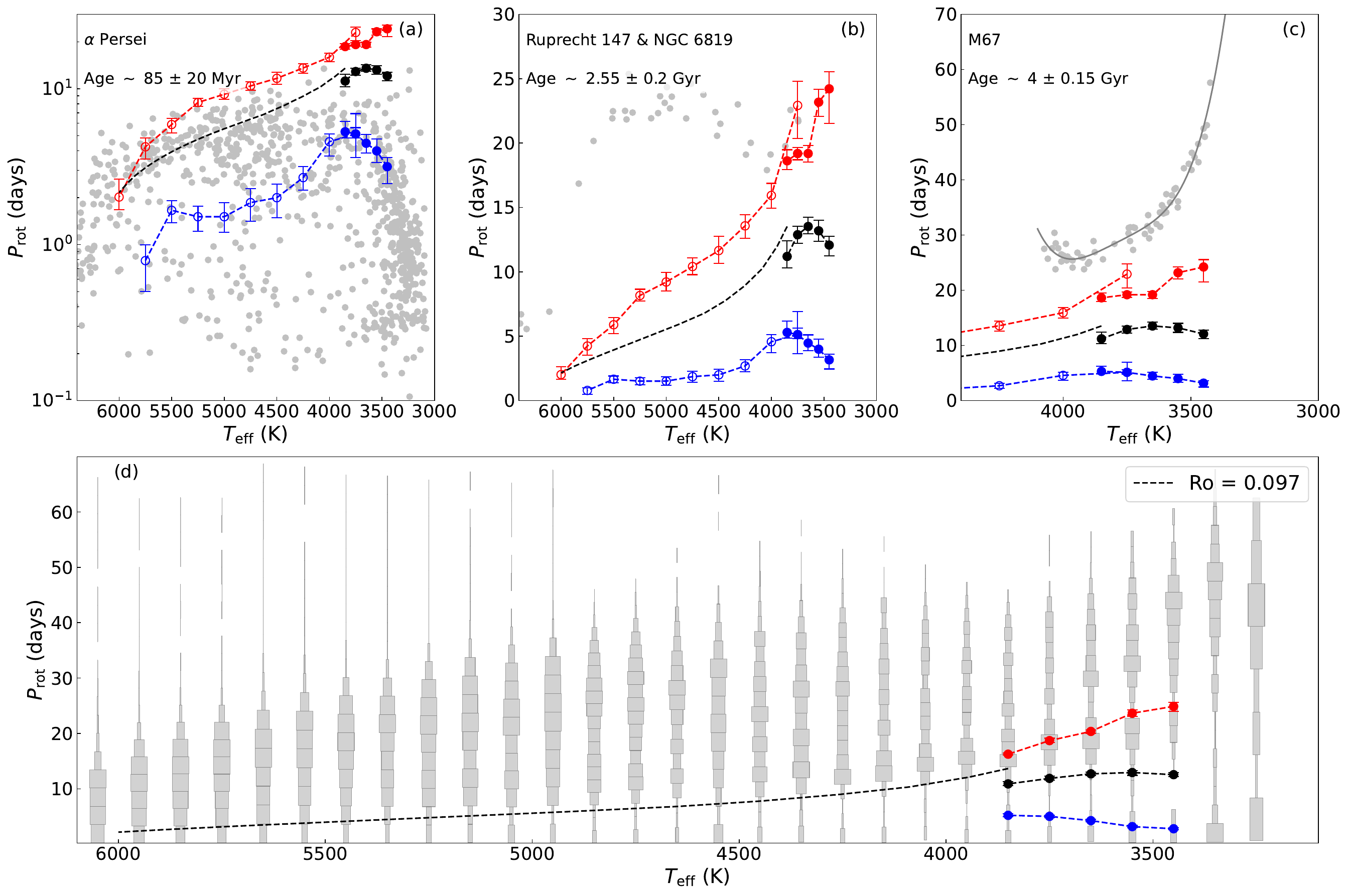}

\caption{$T_{\rm{eff}}$--$P_{\rm{rot}}$ diagrams. Circles with different colors represent different knee points of $R_{\rm{HK}}^{'}$--$P_{\rm{rot}}$ across various $T_{\rm{eff}}$ bins. Blue circles are the first knee points that mark the transition from the ``C'' sequence to the ``Gap'' region. Black circles are the second knee positions where coupling efficiency changes significantly within the ``Gap'' region. Red points represent the third knee points, as the transition between the ``Gap'' region and the ``I'' sequence. Open circles represent results from \cite{2025A&A...699A.251Y}, while filled circles denote our results. Black dashed line in each panel represents the model from \cite{2010ApJ...721..675B} corresponding to Ro = 0.097. Panel (a) displays data for open cluster $\alpha$ Persei \citep{2023AJ....166...14B}. Panel (b) shows the open clusters Ruprecht 147 and NGC 6819 \citep{2020ApJ...904..140C}, with $(V-K_{s})_{0}$ colors converted to $T_{\rm{eff}}$ using \cite{2010ApJ...721..675B} calibrations. Panel (c) presents the open cluster M67 \citep{2022ApJ...938..118D}, where the gray curve and gray points represent the fit and underlying data from \cite{2022ApJ...938..118D}. Panel (d): $T_{\rm{eff}}$--$P_{\rm{rot}}$ diagram of \emph{Kepler} targets from \cite{2014ApJS..211...24M}.}
\label{cluster.fig}
\end{figure*}

\section{Results: a new rotation--activity relation}
\label{sec:relation}
There are 2,884 LAMOST M dwarfs and 242 DESI M dwarfs with both $R_{\rm{HK}}^{'}$  and $P_{\rm{rot}}$ measurements. The final results, including the $S_{\rm{MW}}$, $R_{\rm{HK}}^{'}$, $P_{\rm{rot}}$, $\tau_{\rm{c}}$ together with stellar parameters are given in Table \ref{tab:table1} and Table \ref{tab:table2}. 

The canonical rotation--activity relation \citep{2011ApJ...743...48W} is characterized by two distinct regions. In the saturated region, the activity level remains constant, which may be due to the saturation of magnetic flux (or filling factor) \citep[e.g.,][]{2009ApJ...692..538R}. In the decay region, the activity level depends on Ro, suggesting that faster rotators generally exhibit stronger magnetic activity. 

However, such a picture may not be universal. Recent studies based on X-ray observations have reported a slight decreasing trend (toward slower rotation) in the saturated region \citep[e.g.,][]{2014ApJ...794..144R, 2021A&A...649A..96J,2024ApJS..273....8H,2025MNRAS.539.1922S}. 
The decay region also exhibits discrepancies among various studies. \cite{2018A&A...618A..48M} proposed an empirical three-part model to describe this region. \cite{2019A&A...628A..41P} found that F-type stars deviate from the canonical rotation--activity relation, confirmed subsequently by \cite{2023ApJS..264...12H, 2024ApJS..273....8H}. \cite{2021ApJ...910..110L} proposed a knee point in the decay region. Most recently, \cite{2024A&A...684A.121F} employed a high-order polynomial to model the overall relation using eROSITA data. 

The rotation--activity relation of our M dwarf sample is shown in Figure \ref{ar_fit_piece.fig} and \ref{ar_all.fig}. Obviously, our rotation--activity relation differs from the classical picture. We suggest that a four‑segment piecewise power‑law model can appropriately describe such relation. Using Markov Chain Monte Carlo sampling implemented through the emcee Python package \citep{2013PASP..125..306F}, the model was determined following
\begin{equation}
\small
\begin{aligned}
\log_{10}(R'_{\text{HK}}) &= 
\begin{cases} 
b, & x < x_0, \\
\begin{aligned}
&a_0(x-x_0) + b,
\end{aligned} & x_0 \leq x < x_1, \\
\begin{aligned}
&a_1(x-x_1) \\
&\quad + a_0(x_1-x_0) + b,
\end{aligned} & x_1 \leq x < x_2, \\
\begin{aligned}
&a_2(x-x_2) \\
&\quad + a_1(x_2-x_1) \\
&\quad + a_0(x_1-x_0) + b,
\end{aligned} & x \geq x_2,
\end{cases} \\
&= 
\begin{cases}
-4.171, & x < -1.543, \\
\begin{aligned}
-0.498 * x - 4.939, \end{aligned} & -1.543 \leq x < -1.013, \\
\begin{aligned}
-1.941 * x - 6.401, \end{aligned} & -1.013 \leq x < -0.88, \\
\begin{aligned}
-0.387 * x - 5.034, \end{aligned} & x \geq -0.88,

\end{cases}
\end{aligned}
\label{eq:piecewise_linear}
\end{equation}

where x $\equiv$ $\text{log}_{10}(\text{Ro})$ and b is the saturated $\text{log}_{10}(R_{\text{HK}}^{'})$. Such a four-part piecewise model can also be resolved through a binning process of the rotation--activity relation along the Ro axis. We adopted a binning approach, dividing the Ro axis into intervals of 0.1 width with 0.05 steps. 
For each bin, we calculated the median value of $\text{log}_{10}(R_{\text{HK}}^{'})$, with uncertainties derived as half of the 16th–84th percentile range divided by the square root of the number of stars in the bin. Then we performed the fitting using the binned data.
The fitting results of both methods are summarized in Table \ref{tab:table3}. Posterior probability distributions of the parameters are given in Figure \ref{post.fig} and Figure \ref{post_binned.fig}. Obviously, the fitting results are consistent (Figure \ref{ar_fit_piece.fig} and Table \ref{tab:table3}). 

\begin{table*}
  \centering
  \caption{Best-fit parameters of $R_{\rm{HK}}^{'}$--Ro relation.}
  \label{tab:table3}
  \setlength{\tabcolsep}{4pt}
  \begin{tabular}{c|ccccccc}
    \hline
    Method & \multicolumn{6}{c}{Parameters} \\
    \hline
    \multirow{2}{*}{All points fitting} 
    & $x_{0}$ & $x_{1}$ & $x_{2}$ & $a_{0}$ & $a_{1}$ & $a_{2}$ & $b$ \\
    \cline{2-8}
    & $-1.543^{+0.038}_{-0.032}$ & 
    $-1.013^{+0.019}_{-0.032}$ & $-0.88^{+0.012}_{-0.011}$ & $-0.498^{+0.053}_{-0.055}$ & $-1.941^{+0.289}_{-0.317}$ & $-0.387^{+0.047}_{-0.047}$ & 
    $-4.171^{+0.007}_{-0.007}$
    \\
    \hline
    \multirow{2}{*}{Binned median fitting} 
    & $x_{0}$ & $x_{1}$ & $x_{2}$ & $a_{0}$ & $a_{1}$ & $a_{2}$ & $b$ \\
    \cline{2-8}
    & $-1.67^{+0.045}_{-0.052}$ & $-1.067^{+0.017}_{-0.015}$ & $-0.88^{+0.008}_{-0.008}$ & $-0.328^{+0.035}_{-0.038}$ & $-1.873^{+0.134}_{-0.173}$ & $-0.343^{+0.026}_{-0.027}$ &
    $-4.168^{+0.006}_{-0.005}$
    \\
    \hline
  \end{tabular}

\end{table*}

As with the classical rotation--activity relation, we identify a saturation region where $R_{\rm{HK}}^{'}$ remains constant (Figure \ref{ar_fit_piece.fig}). The first knee point is $\text{log}_{10}(\text{Ro}) = -1.543^{+0.038}_{-0.032}$ for the all points fitting and $\text{log}_{10}(\text{Ro}) = -1.67^{+0.045}_{-0.052}$ for the binned median fitting, both of which are similar to the knee points reported by \cite{2018A&A...618A..48M} and \cite{2025A&A...699A.251Y}. We did not identify a supersaturation region in our relation.

Beyond the saturation region lies the decay region. Our analysis shows that the decay region comprises some distinct parts (Figure \ref{ar_fit_piece.fig}).
We identify three decay regions: (1) A gradual decay between $\text{log}_{10}(\text{Ro}) = -1.543^{+0.038}_{-0.032}$ and $\text{log}_{10}(\text{Ro}) =-1.013^{+0.019}_{-0.032}$. (2) A steep decline from $\text{log}_{10}(\text{Ro}) = -1.013^{+0.019}_{-0.032}$ to $\text{log}_{10}(\text{Ro}) = -0.88^{+0.012}_{-0.011}$. (3) A moderate decay at larger Ro values, i.e., $\text{log}_{10}(\text{Ro}) > -0.88^{+0.012}_{-0.011}$. The last knee point $\text{log}_{10}(\text{Ro}) = -0.88^{+0.012}_{-0.011}$ is consistent with the results of previous studies \citep{2018A&A...618A..48M, 2025A&A...699A.251Y}. For the binned median fitting process, similar results were presented. In addition, we also repeated the fitting processes in the linear-log scale and got similar results (Figure \ref{ar_fit_piece_linear.fig}). 

We further tried models with fewer segments, i.e., two or three segments, to describe the rotation--activity relation. The fitting results are given in Figure \ref{ar_fit_various_models.fig}. To test the statistical significance of the four-segment model, we carried out partial F-tests for nested model comparison to the models. The comparison between the three- and two-segment models yields $P_{\rm{value}}=3.33\times10^{-16}$, indicating that the three-segment model fits the data significantly better than the two-segment model. Meanwhile, comparing the four- versus three-segment models gives $P_{\rm{value}}=0.02$, which suggests that the four-segment model offers a further statistical improvement over the three-segment model. In addition, we used the piecewise regression python package given by \cite{Pilgrim2021} to fit the rotation--activity relation using binned median data (Figure \ref{ar_fit_various_P21.fig}). Despite a slight slope in the first segment, the fitted knee points ($\text{log}_{10}(\text{Ro}) = -1.556\pm0.056$, $-1.057\pm0.016$, and $-0.88\pm0.016$) are consistent with our earlier results. Although the F-test indicates that a four-segment model provides a statistically better fit than simpler alternatives, a three-segment model may also describe activity--rotation relation well (Figure \ref{ar_fit_various_models.fig}). Further observations are required to confirm the four-segment model.

\section{Discussion}

\subsection{Linking rotation--activity relation to core-envelope coupling process}
\label{sec:gyro}

Gyrochronology describes the spin-down history of late-type stars with different masses, which is caused by magnetic braking \citep{2003ApJ...586..464B}. Generally, the gyrochronology consists of three phases: (1) extremely fast-rotating stars reside on the ``C'' sequence, where the turbulent dynamo operates; (2) slow rotators lie on the ``I'' sequence, with magnetic field generation primarily driven by shear between the convective envelope and radiative core (i.e., interface dynamo); (3) some stars fall in the ``Gap'' region, representing a transitional phase from ``C'' to ``I'' sequence.

The rotation--activity relation from \cite{2025A&A...699A.251Y} exhibits four segments, separated by three knee points at $\text{log}_{10}(\text{Ro}) \sim -1.65$,  $-0.82$ and $-0.15$. Based on the nonlinear model from \cite{2010ApJ...722..222B}, which described the evolution of Ro with stellar ages, \cite{2025A&A...699A.251Y} mapped the first three segments to corresponding phases of gyrochronology. They associated the fast-rotating region, where activity levels remain constant, with the ``C'' sequence. Following two regions have different activity decay rates, corresponding to the ``Gap'' and ``I'' sequence, respectively.

\begin{figure*}
\centering
\includegraphics[width=0.95\textwidth]{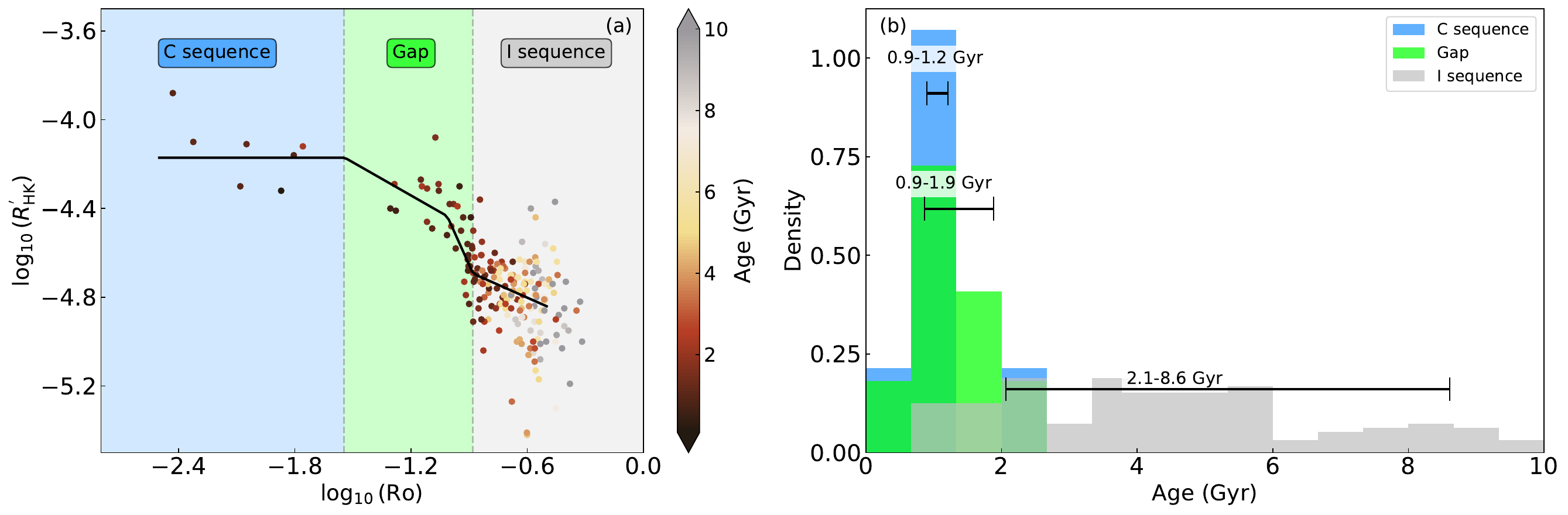}
\caption{Panel (a): $R_{\rm{HK}}^{'}$--Ro relation of stars with measured ages. The colorbar represents stellar ages. Black line is best fit model corresponding to the all points fitting method. Panel (b): Age distributions of stars in each sequence (i.e., ``C'' sequence, ``Gap'', and ``I'' sequence).}
\label{ar_age.fig}
\end{figure*}

As is well known, stars on the ``C'' sequence have largely decoupled inner radiative cores and outer convective envelopes, while those on the ``I'' sequence exhibit full core–envelope coupling \citep{2007ApJ...669.1167B}. The intermediate ``Gap'' sequence therefore represents the process of core–envelope coupling. As stars evolve off the ``C'' sequence, progressive core-envelope coupling reduces the differential rotation between the core and envelope, leading to a decline in activity levels.

Our results reveal a new rotation--activity relation comprising four regions. In comparison with \cite{2025A&A...699A.251Y}, the small-Ro region ($\log_{10}(\text{Ro}) \lesssim -1.543$) aligns with the ``C" sequence, while the large-Ro region ($\log_{10}(\text{Ro}) \gtrsim -0.88$) corresponds to the ``I" sequence. The two intermediate regions represent the transitional ``Gap" sequence. The discovery of this subdivided ``Gap" reveals a variable activity decay rate, thereby implying a non-linear evolution of the underlying magnetic field strength across this regime.

The model of \cite{2020A&A...636A..76S} assumes a constant core-envelope coupling rate. In contrast, the fine structure of the ``Gap" region in our new rotation--activity relation indicates that this rate varies, implying distinct physical processes in the stellar interior. We speculate that the observed change in the activity decay rate is driven by evolving coupling efficiency. Initially low coupling efficiency leads to a slow decay of magnetic activity. As the efficiency increases, the resulting reduction in differential rotation strongly weakens the magnetic field, thereby accelerating the activity decline. Simultaneously, the angular momentum transferred from the core to the envelope (nearly) balances the loss from magnetic braking, producing the well-known stalling of spin-down \citep[e.g.,][]{2018ApJ...862...33A, 2020A&A...636A..76S, 2020ApJ...904..140C}.

These features are also clearly resolved in the $R_{\rm{HK}}^{'}$--$P_{\rm{rot}}$ relation. We divided the sample into different $T_{\rm{eff}}$ bins, in which the convective turnover timescale $\tau_c$ can be considered as constants. For each bin, we also applied a four-part piecewise model, which reproduces the same conclusions (Figure \ref{teff_bin.fig}).
It can be seen that cooler stars exhibit a broader period range within the ``Gap" region. This can be attributed to their thicker convective envelopes, which consequently require a longer timescale for core-envelope coupling.
Both the fit using all data points and the fit to the binned medians yield consistent results (Figure \ref{comparison.fig}).

\subsection{Application to stars with age measurements}

We constructed a period--color diagram for open clusters and field stars of varying ages (Figure \ref{cluster.fig}), overlaying these knee points derived from different $T_{\rm eff}$ subsamples to separate distinct sequences in gyrochronology. The branches defined by the first and third knee points for each subsample mark the times at which different types of stars enter the ``C'' and ``I'' sequences, respectively. 
The second knee points mark the time at which the core-envelope coupling efficiency increases significantly and notable stellar spin-down stalling is expected to start.

In panel (a) of Figure \ref{cluster.fig}, the rotational periods of the third knee points (red dots) are longer than those of members with $T_{\rm{eff}}$ $\lesssim$ 5400 K in the young open cluster $\alpha$ Persei, suggesting these stars haven't yet entered the ``I'' sequence. Stars above the first knee-point branch (blue dots) have left the ``C'' sequence. 
In panel (b), the period sequence of stars cooler than 4000 K in Ruprecht 147 or NGC 6819 lies slightly above the third knee-point branch, suggesting they may have recently entered the ``I" sequence. This explanation is consistent with the finding of \cite{2020ApJ...904..140C} that these stars have experienced a spin-down stalling lasting at least 1.3 Gyr. We note, however, that this comparison relies on an accurate determination of the knee-points. Future observations of more M dwarfs will help to refine the location of this branch and confirm whether these stars have definitively transitioned into the ``I" sequence. 
The slight discrepancy between our results and those of \cite{2025A&A...699A.251Y} may also be attributed to the difference in sample size, with approximately 3000 stars in our sample versus 306 in theirs.
In panel (c), all stars in M67 are located above the third knee-point branch, confirming their entry into the ``I'' sequence. 

Panel (d) displays the double-peaked period distribution of \emph{Kepler} targets across different effective temperatures \citep{2014ApJS..211...24M}. The clustering of stars at the low-period peak has been attributed to the core-envelope coupling process that causes spin-down stalling \citep{2022AJ....164..251L}. Notably, our second and third knee-point branches roughly encompass the low-period peak. This strongly indicates that the second knee-point marks the onset of significant spin-down stalling, implying a sharp increase in core-envelope coupling efficiency. We also plot the model from \cite{2010ApJ...721..675B} corresponding to Ro $=$ 0.096 (i.e., the second knee point). Its close match with the lower period envelope further supports our explanation.

In addition, we found that the periods of the first and second knee points gradually decrease toward lower temperatures. This implies that cooler stars depart from the ``C" sequence and enter the core-envelope coupling phase at shorter rotation periods. However, this trend could also be caused by the limited size of current sample. A larger sample including both partially and fully convective M dwarfs is required to test this hypothesis.

Finally, to establish a quantitative link between the rotation--activity relation and gyrochronology, we cross-matched our sample with the catalog of \citet{2021AJ....161..189L}, which provides gyro-kinematic ages for approximately 30,000 \emph{Kepler} stars.
Figure \ref{ar_age.fig} shows age histograms for different regions of the rotation--activity relation, with the 16th to 84th percentile age ranges marked in Panel (b).
Although the sample size with age measurements is limited, the three regions exhibit roughly distinct age distributions. For example, stars with large Ro values and low activity levels (i.e., on the ``I'' sequence) are much older than those in the other two regions. 
This provides direct observational support for the connection between the rotation--activity relation and stellar gyrochronology.

\section{Summary}
\label{sec:sum}

In this work, by analyzing a large sample of M dwarfs from the LAMOST and DESI spectroscopic surveys, we conducted a new study of the rotation--activity relation and its connection to gyrochronology. We suggest that the new relation can be divided into three parts, corresponding to the three sequences in gyrochronology: \\
(1) The saturated region with small Ro values, where magnetic activity keeps constant, corresponds to the ``C'' sequence. \\
(2) The intermediate region, where the activity decay rate varies and thus consists of two parts, corresponds to the ``Gap'' sequence. \\
(3) The slow-rotator region with large Ro values, where the activity gradually declines, corresponds to the ``I'' sequence.

The variable activity decay rate in the intermediate region, which was discovered for the first time, points to a varying core-envelope coupling efficiency. As the efficiency increases, the resulting sharp reduction in differential rotation substantially weakens the magnetic field, leading to a rapid decline in magnetic activity.
Simultaneously, angular momentum is rapidly transferred from the core to the envelope, roughly balancing the loss due to magnetic breaking. This process maintains nearly constant rotation periods, which is in good agreement with the observed spin-down stalling in open clusters \citep[e.g.,][]{2018ApJ...862...33A,  2020ApJ...904..140C} and the low-period peak identified among \emph{Kepler} stars \citep{2014ApJS..211...24M}.
The varying core-envelope coupling efficiency departs from the constant value assumed by models like
\cite{2020A&A...636A..76S}.
As such, our results provide new constraints on the physical processes operating within stellar interiors.

\begin{acknowledgements}
We thank the referee for the comprehensive suggestions, which have significantly improved the manuscript. We thank Dr. F. Spada for useful discussions about stellar core-envelope coupling process.
The Guoshoujing Telescope (the Large Sky Area Multi-Object Fiber Spectroscopic Telescope LAMOST) is a National Major Scientific Project built by the Chinese Academy of Sciences. Funding for the project has been provided by the National Development and Reform Commission. LAMOST is operated and managed by the National Astronomical Observatories, Chinese Academy of Sciences. 
This work was supported by National Natural Science Foundation of China (NSFC) under grant Nos. 12588202/12273057/11833002/12090042, the National Key Research and Development Program of China (NKRDPC) under grant number 2023YFA1607901, the Strategic Priority Program of the Chinese Academy of Sciences under grant number XDB1160302, and science research grants from the China Manned Space Project. J.F.L acknowledges the support from the New Cornerstone Science Foundation through the New Cornerstone Investigator Program and the XPLORER PRIZE.
\end{acknowledgements}

\bibliography{main}{}

@ARTICLE{1988ApJ...333..236K,
       author = {{Kawaler}, Steven D.},
        title = "{Angular Momentum Loss in Low-Mass Stars}",
      journal = {\apj},
         year = 1988,
        month = oct,
       volume = {333},
        pages = {236},
          doi = {10.1086/166740},
       adsurl = {https://ui.adsabs.harvard.edu/abs/1988ApJ...333..236K}
}

@ARTICLE{2003ApJ...586..464B,
       author = {{Barnes}, Sydney A.},
        title = "{On the Rotational Evolution of Solar- and Late-Type Stars, Its Magnetic Origins, and the Possibility of Stellar Gyrochronology}",
      journal = {\apj},
         year = 2003,
        month = mar,
       volume = {586},
       number = {1},
        pages = {464-479},
          doi = {10.1086/367639},
archivePrefix = {arXiv},
       eprint = {astro-ph/0303631},
 primaryClass = {astro-ph},
       adsurl = {https://ui.adsabs.harvard.edu/abs/2003ApJ...586..464B}
}

@ARTICLE{2016ApJ...823...16B,
       author = {{Barnes}, Sydney A. and {Weingrill}, Joerg and {Fritzewski}, Dario and {Strassmeier}, Klaus G. and {Platais}, Imants},
        title = "{Rotation Periods for Cool Stars in the 4 Gyr old Open Cluster M67, The Solar-Stellar Connection, and the Applicability of Gyrochronology to at least Solar Age}",
      journal = {\apj},
         year = 2016,
        month = may,
       volume = {823},
       number = {1},
          eid = {16},
        pages = {16},
          doi = {10.3847/0004-637X/823/1/16},
archivePrefix = {arXiv},
       eprint = {1603.09179},
 primaryClass = {astro-ph.SR},
       adsurl = {https://ui.adsabs.harvard.edu/abs/2016ApJ...823...16B}
}

@ARTICLE{2009ApJ...695..679M,
       author = {{Meibom}, S{\o}ren and {Mathieu}, Robert D. and {Stassun}, Keivan G.},
        title = "{Stellar Rotation in M35: Mass-Period Relations, Spin-Down Rates, and Gyrochronology}",
      journal = {\apj},
         year = 2009,
        month = apr,
       volume = {695},
       number = {1},
        pages = {679-694},
          doi = {10.1088/0004-637X/695/1/679},
archivePrefix = {arXiv},
       eprint = {0805.1040},
 primaryClass = {astro-ph},
       adsurl = {https://ui.adsabs.harvard.edu/abs/2009ApJ...695..679M}
}

@ARTICLE{2011ApJ...733..115M,
       author = {{Meibom}, S{\o}ren and {Mathieu}, Robert D. and {Stassun}, Keivan G. and {Liebesny}, Paul and {Saar}, Steven H.},
        title = "{The Color-period Diagram and Stellar Rotational Evolution{\textemdash}New Rotation Period Measurements in the Open Cluster M34}",
      journal = {\apj},
         year = 2011,
        month = jun,
       volume = {733},
       number = {2},
          eid = {115},
        pages = {115},
          doi = {10.1088/0004-637X/733/2/115},
archivePrefix = {arXiv},
       eprint = {1103.5171},
 primaryClass = {astro-ph.SR},
       adsurl = {https://ui.adsabs.harvard.edu/abs/2011ApJ...733..115M}
}

@ARTICLE{2015Natur.517..589M,
       author = {{Meibom}, S{\o}ren and {Barnes}, Sydney A. and {Platais}, Imants and {Gilliland}, Ronald L. and {Latham}, David W. and {Mathieu}, Robert D.},
        title = "{A spin-down clock for cool stars from observations of a 2.5-billion-year-old cluster}",
      journal = {\nat},
         year = 2015,
        month = jan,
       volume = {517},
       number = {7536},
        pages = {589-591},
          doi = {10.1038/nature14118},
archivePrefix = {arXiv},
       eprint = {1501.05651},
 primaryClass = {astro-ph.SR},
       adsurl = {https://ui.adsabs.harvard.edu/abs/2015Natur.517..589M}
}

@ARTICLE{2020ApJ...904..140C,
       author = {{Curtis}, Jason Lee and {Ag{\"u}eros}, Marcel A. and {Matt}, Sean P. and {Covey}, Kevin R. and {Douglas}, Stephanie T. and {Angus}, Ruth and {Saar}, Steven H. and {Cody}, Ann Marie and {Vanderburg}, Andrew and {Law}, Nicholas M. and {Kraus}, Adam L. and {Latham}, David W. and {Baranec}, Christoph and {Riddle}, Reed and {Ziegler}, Carl and {Lund}, Mikkel N. and {Torres}, Guillermo and {Meibom}, S{\o}ren and {Aguirre}, Victor Silva and {Wright}, Jason T.},
        title = "{When Do Stalled Stars Resume Spinning Down? Advancing Gyrochronology with Ruprecht 147}",
      journal = {\apj},
         year = 2020,
        month = dec,
       volume = {904},
       number = {2},
          eid = {140},
        pages = {140},
          doi = {10.3847/1538-4357/abbf58},
archivePrefix = {arXiv},
       eprint = {2010.02272},
 primaryClass = {astro-ph.SR},
       adsurl = {https://ui.adsabs.harvard.edu/abs/2020ApJ...904..140C}
}

@ARTICLE{2016Natur.529..181V,
       author = {{van Saders}, Jennifer L. and {Ceillier}, Tugdual and {Metcalfe}, Travis S. and {Silva Aguirre}, Victor and {Pinsonneault}, Marc H. and {Garc{\'\i}a}, Rafael A. and {Mathur}, Savita and {Davies}, Guy R.},
        title = "{Weakened magnetic braking as the origin of anomalously rapid rotation in old field stars}",
      journal = {\nat},
         year = 2016,
        month = jan,
       volume = {529},
       number = {7585},
        pages = {181-184},
          doi = {10.1038/nature16168},
archivePrefix = {arXiv},
       eprint = {1601.02631},
 primaryClass = {astro-ph.SR},
       adsurl = {https://ui.adsabs.harvard.edu/abs/2016Natur.529..181V}
}

@ARTICLE{2020A&A...636A..76S,
       author = {{Spada}, F. and {Lanzafame}, A.~C.},
        title = "{Competing effect of wind braking and interior coupling in the rotational evolution of solar-like stars}",
      journal = {\aap},
         year = 2020,
        month = apr,
       volume = {636},
          eid = {A76},
        pages = {A76},
          doi = {10.1051/0004-6361/201936384},
archivePrefix = {arXiv},
       eprint = {1908.00345},
 primaryClass = {astro-ph.SR},
       adsurl = {https://ui.adsabs.harvard.edu/abs/2020A&A...636A..76S}
}

@ARTICLE{2022AJ....164..251L,
       author = {{Lu}, Yuxi Lucy and {Curtis}, Jason L. and {Angus}, Ruth and {David}, Trevor J. and {Hattori}, Soichiro},
        title = "{Bridging the Gap-The Disappearance of the Intermediate Period Gap for Fully Convective Stars, Uncovered by New ZTF Rotation Periods}",
      journal = {\aj},
         year = 2022,
        month = dec,
       volume = {164},
       number = {6},
          eid = {251},
        pages = {251},
          doi = {10.3847/1538-3881/ac9bee},
archivePrefix = {arXiv},
       eprint = {2210.06604},
 primaryClass = {astro-ph.SR},
       adsurl = {https://ui.adsabs.harvard.edu/abs/2022AJ....164..251L}
}

@ARTICLE{2011ApJ...743...48W,
       author = {{Wright}, Nicholas J. and {Drake}, Jeremy J. and {Mamajek}, Eric E. and {Henry}, Gregory W.},
        title = "{The Stellar-activity-Rotation Relationship and the Evolution of Stellar Dynamos}",
      journal = {\apj},
         year = 2011,
        month = dec,
       volume = {743},
       number = {1},
          eid = {48},
        pages = {48},
          doi = {10.1088/0004-637X/743/1/48},
archivePrefix = {arXiv},
       eprint = {1109.4634},
 primaryClass = {astro-ph.SR},
       adsurl = {https://ui.adsabs.harvard.edu/abs/2011ApJ...743...48W}
}

@ARTICLE{2014ApJ...794..144R,
       author = {{Reiners}, A. and {Sch{\"u}ssler}, M. and {Passegger}, V.~M.},
        title = "{Generalized Investigation of the Rotation-Activity Relation: Favoring Rotation Period instead of Rossby Number}",
      journal = {\apj},
         year = 2014,
        month = oct,
       volume = {794},
       number = {2},
          eid = {144},
        pages = {144},
          doi = {10.1088/0004-637X/794/2/144},
archivePrefix = {arXiv},
       eprint = {1408.6175},
 primaryClass = {astro-ph.SR},
       adsurl = {https://ui.adsabs.harvard.edu/abs/2014ApJ...794..144R}
}

@ARTICLE{2019A&A...628A..41P,
       author = {{Pizzocaro}, D. and {Stelzer}, B. and {Poretti}, E. and {Raetz}, S. and {Micela}, G. and {Belfiore}, A. and {Marelli}, M. and {Salvetti}, D. and {De Luca}, A.},
        title = "{Activity and rotation of the X-ray emitting Kepler stars}",
      journal = {\aap},
         year = 2019,
        month = aug,
       volume = {628},
          eid = {A41},
        pages = {A41},
          doi = {10.1051/0004-6361/201731674},
archivePrefix = {arXiv},
       eprint = {1906.05587},
 primaryClass = {astro-ph.SR},
       adsurl = {https://ui.adsabs.harvard.edu/abs/2019A&A...628A..41P}
}

@ARTICLE{2024ApJS..273....8H,
       author = {{Han}, Henggeng and {Wang}, Song and {Zheng}, Chuanjie and {Li}, Xue and {Xiao}, Kai and {Liu}, Jifeng},
        title = "{Stellar X-Ray Activity and Habitability Revealed by the ROSAT Sky Survey}",
      journal = {\apjs},
         year = 2024,
        month = jul,
       volume = {273},
       number = {1},
          eid = {8},
        pages = {8},
          doi = {10.3847/1538-4365/ad4b17},
archivePrefix = {arXiv},
       eprint = {2405.02863},
 primaryClass = {astro-ph.SR},
       adsurl = {https://ui.adsabs.harvard.edu/abs/2024ApJS..273....8H}
}

@ARTICLE{2018A&A...618A..48M,
       author = {{Mittag}, M. and {Schmitt}, J.~H.~M.~M. and {Schr{\"o}der}, K. -P.},
        title = "{Revisiting the connection between magnetic activity, rotation period, and convective turnover time for main-sequence stars}",
      journal = {\aap},
         year = 2018,
        month = oct,
       volume = {618},
          eid = {A48},
        pages = {A48},
          doi = {10.1051/0004-6361/201833498},
archivePrefix = {arXiv},
       eprint = {1807.05825},
 primaryClass = {astro-ph.SR},
       adsurl = {https://ui.adsabs.harvard.edu/abs/2018A&A...618A..48M}
}

@ARTICLE{2003ApJ...586L.145B,
       author = {{Barnes}, Sydney A.},
        title = "{A Connection between the Morphology of the X-Ray Emission and Rotation for Solar-Type Stars in Open Clusters}",
      journal = {\apjl},
         year = 2003,
        month = apr,
       volume = {586},
       number = {2},
        pages = {L145-L147},
          doi = {10.1086/374681},
archivePrefix = {arXiv},
       eprint = {astro-ph/0303632},
 primaryClass = {astro-ph},
       adsurl = {https://ui.adsabs.harvard.edu/abs/2003ApJ...586L.145B}
}

@ARTICLE{2025A&A...699A.251Y,
       author = {{Yang}, Huiqin and {Liu}, Jifeng and {Soria}, Roberto and {Spada}, Federico and {Wang}, Song and {Fang}, Xiangsong and {Li}, Xue},
        title = "{Four ages of rotating stars in the rotation{\textendash}activity relationship and gyrochronology}",
      journal = {\aap},
         year = 2025,
        month = jul,
       volume = {699},
          eid = {A251},
        pages = {A251},
          doi = {10.1051/0004-6361/202554379},
archivePrefix = {arXiv},
       eprint = {2506.07383},
 primaryClass = {astro-ph.SR},
       adsurl = {https://ui.adsabs.harvard.edu/abs/2025A&A...699A.251Y}
}

@ARTICLE{2012RAA....12.1197C,
       author = {{Cui}, Xiang-Qun and {Zhao}, Yong-Heng and {Chu}, Yao-Quan and {Li}, Guo-Ping and {Li}, Qi and {Zhang}, Li-Ping and {Su}, Hong-Jun and {Yao}, Zheng-Qiu and {Wang}, Ya-Nan and {Xing}, Xiao-Zheng and {Li}, Xin-Nan and {Zhu}, Yong-Tian and {Wang}, Gang and {Gu}, Bo-Zhong and {Luo}, A. -Li and {Xu}, Xin-Qi and {Zhang}, Zhen-Chao and {Liu}, Gen-Rong and {Zhang}, Hao-Tong and {Yang}, De-Hua and {Cao}, Shu-Yun and {Chen}, Hai-Yuan and {Chen}, Jian-Jun and {Chen}, Kun-Xin and {Chen}, Ying and {Chu}, Jia-Ru and {Feng}, Lei and {Gong}, Xue-Fei and {Hou}, Yong-Hui and {Hu}, Hong-Zhuan and {Hu}, Ning-Sheng and {Hu}, Zhong-Wen and {Jia}, Lei and {Jiang}, Fang-Hua and {Jiang}, Xiang and {Jiang}, Zi-Bo and {Jin}, Ge and {Li}, Ai-Hua and {Li}, Yan and {Li}, Ye-Ping and {Liu}, Guan-Qun and {Liu}, Zhi-Gang and {Lu}, Wen-Zhi and {Mao}, Yin-Dun and {Men}, Li and {Qi}, Yong-Jun and {Qi}, Zhao-Xiang and {Shi}, Huo-Ming and {Tang}, Zheng-Hong and {Tao}, Qing-Sheng and {Wang}, Da-Qi and {Wang}, Dan and {Wang}, Guo-Min and {Wang}, Hai and {Wang}, Jia-Ning and {Wang}, Jian and {Wang}, Jian-Ling and {Wang}, Jian-Ping and {Wang}, Lei and {Wang}, Shu-Qing and {Wang}, You and {Wang}, Yue-Fei and {Xu}, Ling-Zhe and {Xu}, Yan and {Yang}, Shi-Hai and {Yu}, Yong and {Yuan}, Hui and {Yuan}, Xiang-Yan and {Zhai}, Chao and {Zhang}, Jing and {Zhang}, Yan-Xia and {Zhang}, Yong and {Zhao}, Ming and {Zhou}, Fang and {Zhou}, Guo-Hua and {Zhu}, Jie and {Zou}, Si-Cheng},
        title = "{The Large Sky Area Multi-Object Fiber Spectroscopic Telescope (LAMOST)}",
      journal = {Research in Astronomy and Astrophysics},
         year = 2012,
        month = sep,
       volume = {12},
       number = {9},
        pages = {1197-1242},
          doi = {10.1088/1674-4527/12/9/003},
       adsurl = {https://ui.adsabs.harvard.edu/abs/2012RAA....12.1197C}
}

@ARTICLE{2015RAA....15.1095L,
       author = {{Luo}, A. -Li and {Zhao}, Yong-Heng and {Zhao}, Gang and {Deng}, Li-Cai and {Liu}, Xiao-Wei and {Jing}, Yi-Peng and {Wang}, Gang and {Zhang}, Hao-Tong and {Shi}, Jian-Rong and {Cui}, Xiang-Qun and {Chu}, Yao-Quan and {Li}, Guo-Ping and {Bai}, Zhong-Rui and {Wu}, Yue and {Cai}, Yan and {Cao}, Shu-Yun and {Cao}, Zi-Huang and {Carlin}, Jeffrey L. and {Chen}, Hai-Yuan and {Chen}, Jian-Jun and {Chen}, Kun-Xin and {Chen}, Li and {Chen}, Xue-Lei and {Chen}, Xiao-Yan and {Chen}, Ying and {Christlieb}, Norbert and {Chu}, Jia-Ru and {Cui}, Chen-Zhou and {Dong}, Yi-Qiao and {Du}, Bing and {Fan}, Dong-Wei and {Feng}, Lei and {Fu}, Jian-Ning and {Gao}, Peng and {Gong}, Xue-Fei and {Gu}, Bo-Zhong and {Guo}, Yan-Xin and {Han}, Zhan-Wen and {He}, Bo-Liang and {Hou}, Jin-Liang and {Hou}, Yong-Hui and {Hou}, Wen and {Hu}, Hong-Zhuan and {Hu}, Ning-Sheng and {Hu}, Zhong-Wen and {Huo}, Zhi-Ying and {Jia}, Lei and {Jiang}, Fang-Hua and {Jiang}, Xiang and {Jiang}, Zhi-Bo and {Jin}, Ge and {Kong}, Xiao and {Kong}, Xu and {Lei}, Ya-Juan and {Li}, Ai-Hua and {Li}, Chang-Hua and {Li}, Guang-Wei and {Li}, Hai-Ning and {Li}, Jian and {Li}, Qi and {Li}, Shuang and {Li}, Sha-Sha and {Li}, Xin-Nan and {Li}, Yan and {Li}, Yin-Bi and {Li}, Ye-Ping and {Liang}, Yuan and {Lin}, Chien-Cheng and {Liu}, Chao and {Liu}, Gen-Rong and {Liu}, Guan-Qun and {Liu}, Zhi-Gang and {Lu}, Wen-Zhi and {Luo}, Yu and {Mao}, Yin-Dun and {Newberg}, Heidi and {Ni}, Ji-Jun and {Qi}, Zhao-Xiang and {Qi}, Yong-Jun and {Shen}, Shi-Yin and {Shi}, Huo-Ming and {Song}, Jing and {Song}, Yi-Han and {Su}, Ding-Qiang and {Su}, Hong-Jun and {Tang}, Zheng-Hong and {Tao}, Qing-Sheng and {Tian}, Yuan and {Wang}, Dan and {Wang}, Da-Qi and {Wang}, Feng-Fei and {Wang}, Guo-Min and {Wang}, Hai and {Wang}, Hong-Chi and {Wang}, Jian and {Wang}, Jia-Ning and {Wang}, Jian-Ling and {Wang}, Jian-Ping and {Wang}, Jun-Xian and {Wang}, Lei and {Wang}, Meng-Xin and {Wang}, Shou-Guan and {Wang}, Shu-Qing and {Wang}, Xia and {Wang}, Ya-Nan and {Wang}, You and {Wang}, Yue-Fei and {Wang}, You-Fen and {Wei}, Peng and {Wei}, Ming-Zhi and {Wu}, Hong and {Wu}, Ke-Fei and {Wu}, Xue-Bing and {Wu}, Yu-Zhong and {Xing}, Xiao-Zheng and {Xu}, Ling-Zhe and {Xu}, Xin-Qi and {Xu}, Yan and {Yan}, Tai-Sheng and {Yang}, De-Hua and {Yang}, Hai-Feng and {Yang}, Hui-Qin and {Yang}, Ming and {Yao}, Zheng-Qiu and {Yu}, Yong and {Yuan}, Hui and {Yuan}, Hai-Bo and {Yuan}, Hai-Long and {Yuan}, Wei-Min and {Zhai}, Chao and {Zhang}, En-Peng and {Zhang}, Hua-Wei and {Zhang}, Jian-Nan and {Zhang}, Li-Pin and {Zhang}, Wei and {Zhang}, Yong and {Zhang}, Yan-Xia and {Zhang}, Zheng-Chao and {Zhao}, Ming and {Zhou}, Fang and {Zhou}, Xu and {Zhu}, Jie and {Zhu}, Yong-Tian and {Zou}, Si-Cheng and {Zuo}, Fang},
        title = "{The first data release (DR1) of the LAMOST regular survey}",
      journal = {Research in Astronomy and Astrophysics},
         year = 2015,
        month = aug,
       volume = {15},
       number = {8},
          eid = {1095},
        pages = {1095},
          doi = {10.1088/1674-4527/15/8/002},
archivePrefix = {arXiv},
       eprint = {1505.01570},
 primaryClass = {astro-ph.GA},
       adsurl = {https://ui.adsabs.harvard.edu/abs/2015RAA....15.1095L}
}

@misc{desicollaboration2025datarelease1dark,
      title={Data Release 1 of the Dark Energy Spectroscopic Instrument}, 
      author={DESI Collaboration and M. Abdul-Karim and A. G. Adame and D. Aguado and J. Aguilar and S. Ahlen and S. Alam and G. Aldering and D. M. Alexander and R. Alfarsy and L. Allen and C. Allende Prieto and O. Alves and A. Anand and U. Andrade and E. Armengaud and S. Avila and A. Aviles and H. Awan and S. Bailey and A. Baleato Lizancos and O. Ballester and A. Bault and J. Bautista and S. BenZvi and L. Beraldo e Silva and J. R. Bermejo-Climent and F. Beutler and D. Bianchi and C. Blake and R. Blum and A. S. Bolton and M. Bonici and S. Brieden and A. Brodzeller and D. Brooks and E. Buckley-Geer and E. Burtin and R. Canning and A. Carnero Rosell and A. Carr and P. Carrilho and L. Casas and F. J. Castander and R. Cereskaite and J. L. Cervantes-Cota and E. Chaussidon and J. Chaves-Montero and S. Chen and X. Chen and T. Claybaugh and S. Cole and A. P. Cooper and M. -C. Cousinou and A. Cuceu and T. M. Davis and K. S. Dawson and R. de Belsunce and R. de la Cruz and A. de la Macorra and A. de Mattia and N. Deiosso and J. Della Costa and R. Demina and U. Demirbozan and J. DeRose and A. Dey and B. Dey and J. Ding and Z. Ding and P. Doel and K. Douglass and M. Dowicz and H. Ebina and J. Edelstein and D. J. Eisenstein and W. Elbers and N. Emas and S. Escoffier and P. Fagrelius and X. Fan and K. Fanning and V. A. Fawcett and E. Fernández-García and S. Ferraro and N. Findlay and A. Font-Ribera and J. E. Forero-Romero and D. Forero-Sánchez and C. S. Frenk and B. T. Gänsicke and L. Galbany and J. García-Bellido and C. Garcia-Quintero and L. H. Garrison and E. Gaztañaga and H. Gil-Marín and O. Y. Gnedin and S. Gontcho A Gontcho and A. X. Gonzalez-Morales and V. Gonzalez-Perez and C. Gordon and O. Graur and D. Green and D. Gruen and R. Gsponer and C. Guandalin and G. Gutierrez and J. Guy and C. Hahn and J. J. Han and J. Han and S. He and H. K. Herrera-Alcantar and K. Honscheid and J. Hou and C. Howlett and D. Huterer and V. Iršič and M. Ishak and A. Jacques and J. Jimenez and Y. P. Jing and B. Joachimi and S. Joudaki and R. Joyce and E. Jullo and S. Juneau and N. G. Karaçaylı and T. Karim and R. Kehoe and S. Kent and A. Khederlarian and D. Kirkby and T. Kisner and F. -S. Kitaura and N. Kizhuprakkat and H. Kong and S. E. Koposov and A. Kremin and A. Krolewski and O. Lahav and Y. Lai and C. Lamman and T. -W. Lan and M. Landriau and D. Lang and J. U. Lange and J. Lasker and J. M. Le Goff and L. Le Guillou and A. Leauthaud and M. E. Levi and S. Li and T. S. Li and K. Lodha and M. Lokken and Y. Luo and C. Magneville and M. Manera and C. J. Manser and D. Margala and P. Martini and M. Maus and J. McCullough and P. McDonald and G. E. Medina and L. Medina-Varela and A. Meisner and J. Mena-Fernández and A. Menegas and M. Mezcua and R. Miquel and P. Montero-Camacho and J. Moon and J. Moustakas and A. Muñoz-Gutiérrez and D. Muñoz-Santos and A. D. Myers and J. Myles and S. Nadathur and J. Najita and L. Napolitano and J. A. Newman and F. Nikakhtar and R. Nikutta and G. Niz and H. E. Noriega and N. Padmanabhan and E. Paillas and N. Palanque-Delabrouille and A. Palmese and J. Pan and Z. Pan and D. Parkinson and J. Peacock and W. J. Percival and A. Pérez-Fernández and I. Pérez-Ràfols and P. Peterson and J. Piat and M. M. Pieri and M. Pinon and C. Poppett and A. Porredon and F. Prada and R. Pucha and F. Qin and D. Rabinowitz and A. Raichoor and C. Ramírez-Pérez and S. Ramirez-Solano and M. Rashkovetskyi and C. Ravoux and A. H. Riley and A. Rocher and C. Rockosi and J. Rohlf and A. J. Ross and G. Rossi and R. Ruggeri and V. Ruhlmann-Kleider and C. G. Sabiu and K. Said and A. Saintonge and L. Samushia and E. Sanchez and N. Sanders and C. Saulder and E. F. Schlafly and D. Schlegel and D. Scholte and M. Schubnell and H. Seo and A. Shafieloo and R. Sharples and J. Silber and M. Siudek and A. Smith and D. Sprayberry and J. Suárez-Pérez and J. Swanson and T. Tan and G. Tarlé and P. Taylor and G. Thomas and R. Tojeiro and R. J. Turner and W. Turner and L. A. Ureña-López and R. Vaisakh and M. Valluri and M. Vargas-Magaña and L. Verde and M. Walther and B. Wang and M. S. Wang and W. Wang and B. A. Weaver and N. Weaverdyck and R. H. Wechsler and M. White and M. Wolfson and J. Yang and C. Yèche and S. Youles and J. Yu and S. Yuan and E. A. Zaborowski and P. Zarrouk and H. Zhang and C. Zhao and R. Zhao and Z. Zheng and R. Zhou and H. Zou and S. Zou and Y. Zu},
      year={2025},
      eprint={2503.14745},
      archivePrefix={arXiv},
      primaryClass={astro-ph.CO},
      url={https://arxiv.org/abs/2503.14745}, 
}

@ARTICLE{2020arXiv200507210L,
       author = {{Liu}, Chao and {Fu}, Jianning and {Shi}, Jianrong and {Wu}, Hong and {Han}, Zhanwen and {Chen}, Li and {Dong}, Subo and {Zhao}, Yongheng and {Chen}, Jian-Jun and {Zhang}, Haotong and {Bai}, Zhong-Rui and {Chen}, Xuefei and {Cui}, Wenyuan and {Du}, Bing and {Hsia}, Chih-Hao and {Jiang}, Deng-Kai and {Hou}, Jinliang and {Hou}, Wen and {Li}, Haining and {Li}, Jiao and {Li}, Lifang and {Liu}, Jiaming and {Liu}, Jifeng and {Luo}, A-Li and {Ren}, Juan-Juan and {Tian}, Hai-Jun and {Tian}, Hao and {Wang}, Jia-Xin and {Wu}, Chao-Jian and {Xie}, Ji-Wei and {Yan}, Hong-Liang and {Yang}, Fan and {Yu}, Jincheng and {Zhang}, Bo and {Zhang}, Huawei and {Zhang}, Li-Yun and {Zhang}, Wei and {Zhao}, Gang and {Zhong}, Jing and {Zong}, Weikai and {Zuo}, Fang},
        title = "{LAMOST Medium-Resolution Spectroscopic Survey (LAMOST-MRS): Scientific goals and survey plan}",
      journal = {arXiv e-prints},
         year = 2020,
        month = may,
          eid = {arXiv:2005.07210},
        pages = {arXiv:2005.07210},
          doi = {10.48550/arXiv.2005.07210},
archivePrefix = {arXiv},
       eprint = {2005.07210},
 primaryClass = {astro-ph.SR},
       adsurl = {https://ui.adsabs.harvard.edu/abs/2020arXiv200507210L}
}

@ARTICLE{2016arXiv161100036D,
       author = {{DESI Collaboration} and {Aghamousa}, Amir and {Aguilar}, Jessica and {Ahlen}, Steve and {Alam}, Shadab and {Allen}, Lori E. and {Allende Prieto}, Carlos and {Annis}, James and {Bailey}, Stephen and {Balland}, Christophe and {Ballester}, Otger and {Baltay}, Charles and {Beaufore}, Lucas and {Bebek}, Chris and {Beers}, Timothy C. and {Bell}, Eric F. and {Bernal}, Jos{\'e} Luis and {Besuner}, Robert and {Beutler}, Florian and {Blake}, Chris and {Bleuler}, Hannes and {Blomqvist}, Michael and {Blum}, Robert and {Bolton}, Adam S. and {Briceno}, Cesar and {Brooks}, David and {Brownstein}, Joel R. and {Buckley-Geer}, Elizabeth and {Burden}, Angela and {Burtin}, Etienne and {Busca}, Nicolas G. and {Cahn}, Robert N. and {Cai}, Yan-Chuan and {Cardiel-Sas}, Laia and {Carlberg}, Raymond G. and {Carton}, Pierre-Henri and {Casas}, Ricard and {Castander}, Francisco J. and {Cervantes-Cota}, Jorge L. and {Claybaugh}, Todd M. and {Close}, Madeline and {Coker}, Carl T. and {Cole}, Shaun and {Comparat}, Johan and {Cooper}, Andrew P. and {Cousinou}, M. -C. and {Crocce}, Martin and {Cuby}, Jean-Gabriel and {Cunningham}, Daniel P. and {Davis}, Tamara M. and {Dawson}, Kyle S. and {de la Macorra}, Axel and {De Vicente}, Juan and {Delubac}, Timoth{\'e}e and {Derwent}, Mark and {Dey}, Arjun and {Dhungana}, Govinda and {Ding}, Zhejie and {Doel}, Peter and {Duan}, Yutong T. and {Ealet}, Anne and {Edelstein}, Jerry and {Eftekharzadeh}, Sarah and {Eisenstein}, Daniel J. and {Elliott}, Ann and {Escoffier}, St{\'e}phanie and {Evatt}, Matthew and {Fagrelius}, Parker and {Fan}, Xiaohui and {Fanning}, Kevin and {Farahi}, Arya and {Farihi}, Jay and {Favole}, Ginevra and {Feng}, Yu and {Fernandez}, Enrique and {Findlay}, Joseph R. and {Finkbeiner}, Douglas P. and {Fitzpatrick}, Michael J. and {Flaugher}, Brenna and {Flender}, Samuel and {Font-Ribera}, Andreu and {Forero-Romero}, Jaime E. and {Fosalba}, Pablo and {Frenk}, Carlos S. and {Fumagalli}, Michele and {Gaensicke}, Boris T. and {Gallo}, Giuseppe and {Garcia-Bellido}, Juan and {Gaztanaga}, Enrique and {Pietro Gentile Fusillo}, Nicola and {Gerard}, Terry and {Gershkovich}, Irena and {Giannantonio}, Tommaso and {Gillet}, Denis and {Gonzalez-de-Rivera}, Guillermo and {Gonzalez-Perez}, Violeta and {Gott}, Shelby and {Graur}, Or and {Gutierrez}, Gaston and {Guy}, Julien and {Habib}, Salman and {Heetderks}, Henry and {Heetderks}, Ian and {Heitmann}, Katrin and {Hellwing}, Wojciech A. and {Herrera}, David A. and {Ho}, Shirley and {Holland}, Stephen and {Honscheid}, Klaus and {Huff}, Eric and {Hutchinson}, Timothy A. and {Huterer}, Dragan and {Hwang}, Ho Seong and {Illa Laguna}, Joseph Maria and {Ishikawa}, Yuzo and {Jacobs}, Dianna and {Jeffrey}, Niall and {Jelinsky}, Patrick and {Jennings}, Elise and {Jiang}, Linhua and {Jimenez}, Jorge and {Johnson}, Jennifer and {Joyce}, Richard and {Jullo}, Eric and {Juneau}, St{\'e}phanie and {Kama}, Sami and {Karcher}, Armin and {Karkar}, Sonia and {Kehoe}, Robert and {Kennamer}, Noble and {Kent}, Stephen and {Kilbinger}, Martin and {Kim}, Alex G. and {Kirkby}, David and {Kisner}, Theodore and {Kitanidis}, Ellie and {Kneib}, Jean-Paul and {Koposov}, Sergey and {Kovacs}, Eve and {Koyama}, Kazuya and {Kremin}, Anthony and {Kron}, Richard and {Kronig}, Luzius and {Kueter-Young}, Andrea and {Lacey}, Cedric G. and {Lafever}, Robin and {Lahav}, Ofer and {Lambert}, Andrew and {Lampton}, Michael and {Landriau}, Martin and {Lang}, Dustin and {Lauer}, Tod R. and {Le Goff}, Jean-Marc and {Le Guillou}, Laurent and {Le Van Suu}, Auguste and {Lee}, Jae Hyeon and {Lee}, Su-Jeong and {Leitner}, Daniela and {Lesser}, Michael and {Levi}, Michael E. and {L'Huillier}, Benjamin and {Li}, Baojiu and {Liang}, Ming and {Lin}, Huan and {Linder}, Eric and {Loebman}, Sarah R. and {Luki{\'c}}, Zarija and {Ma}, Jun and {MacCrann}, Niall and {Magneville}, Christophe and {Makarem}, Laleh and {Manera}, Marc and {Manser}, Christopher J. and {Marshall}, Robert and {Martini}, Paul and {Massey}, Richard and {Matheson}, Thomas and {McCauley}, Jeremy and {McDonald}, Patrick and {McGreer}, Ian D. and {Meisner}, Aaron and {Metcalfe}, Nigel and {Miller}, Timothy N. and {Miquel}, Ramon and {Moustakas}, John and {Myers}, Adam and {Naik}, Milind and {Newman}, Jeffrey A. and {Nichol}, Robert C. and {Nicola}, Andrina and {Nicolati da Costa}, Luiz and {Nie}, Jundan and {Niz}, Gustavo and {Norberg}, Peder and {Nord}, Brian and {Norman}, Dara and {Nugent}, Peter and {O'Brien}, Thomas and {Oh}, Minji and {Olsen}, Knut A.~G.},
        title = "{The DESI Experiment Part I: Science,Targeting, and Survey Design}",
      journal = {arXiv e-prints},
         year = 2016,
        month = oct,
          eid = {arXiv:1611.00036},
        pages = {arXiv:1611.00036},
          doi = {10.48550/arXiv.1611.00036},
archivePrefix = {arXiv},
       eprint = {1611.00036},
 primaryClass = {astro-ph.IM},
       adsurl = {https://ui.adsabs.harvard.edu/abs/2016arXiv161100036D}
}

@ARTICLE{2021A&A...649A...1G,
       author = {{Gaia Collaboration} and {Brown}, A.~G.~A. and {Vallenari}, A. and {Prusti}, T. and {de Bruijne}, J.~H.~J. and {Babusiaux}, C. and {Biermann}, M. and {Creevey}, O.~L. and {Evans}, D.~W. and {Eyer}, L. and {Hutton}, A. and {Jansen}, F. and {Jordi}, C. and {Klioner}, S.~A. and {Lammers}, U. and {Lindegren}, L. and {Luri}, X. and {Mignard}, F. and {Panem}, C. and {Pourbaix}, D. and {Randich}, S. and {Sartoretti}, P. and {Soubiran}, C. and {Walton}, N.~A. and {Arenou}, F. and {Bailer-Jones}, C.~A.~L. and {Bastian}, U. and {Cropper}, M. and {Drimmel}, R. and {Katz}, D. and {Lattanzi}, M.~G. and {van Leeuwen}, F. and {Bakker}, J. and {Cacciari}, C. and {Casta{\~n}eda}, J. and {De Angeli}, F. and {Ducourant}, C. and {Fabricius}, C. and {Fouesneau}, M. and {Fr{\'e}mat}, Y. and {Guerra}, R. and {Guerrier}, A. and {Guiraud}, J. and {Jean-Antoine Piccolo}, A. and {Masana}, E. and {Messineo}, R. and {Mowlavi}, N. and {Nicolas}, C. and {Nienartowicz}, K. and {Pailler}, F. and {Panuzzo}, P. and {Riclet}, F. and {Roux}, W. and {Seabroke}, G.~M. and {Sordo}, R. and {Tanga}, P. and {Th{\'e}venin}, F. and {Gracia-Abril}, G. and {Portell}, J. and {Teyssier}, D. and {Altmann}, M. and {Andrae}, R. and {Bellas-Velidis}, I. and {Benson}, K. and {Berthier}, J. and {Blomme}, R. and {Brugaletta}, E. and {Burgess}, P.~W. and {Busso}, G. and {Carry}, B. and {Cellino}, A. and {Cheek}, N. and {Clementini}, G. and {Damerdji}, Y. and {Davidson}, M. and {Delchambre}, L. and {Dell'Oro}, A. and {Fern{\'a}ndez-Hern{\'a}ndez}, J. and {Galluccio}, L. and {Garc{\'\i}a-Lario}, P. and {Garcia-Reinaldos}, M. and {Gonz{\'a}lez-N{\'u}{\~n}ez}, J. and {Gosset}, E. and {Haigron}, R. and {Halbwachs}, J. -L. and {Hambly}, N.~C. and {Harrison}, D.~L. and {Hatzidimitriou}, D. and {Heiter}, U. and {Hern{\'a}ndez}, J. and {Hestroffer}, D. and {Hodgkin}, S.~T. and {Holl}, B. and {Jan{\ss}en}, K. and {Jevardat de Fombelle}, G. and {Jordan}, S. and {Krone-Martins}, A. and {Lanzafame}, A.~C. and {L{\"o}ffler}, W. and {Lorca}, A. and {Manteiga}, M. and {Marchal}, O. and {Marrese}, P.~M. and {Moitinho}, A. and {Mora}, A. and {Muinonen}, K. and {Osborne}, P. and {Pancino}, E. and {Pauwels}, T. and {Petit}, J. -M. and {Recio-Blanco}, A. and {Richards}, P.~J. and {Riello}, M. and {Rimoldini}, L. and {Robin}, A.~C. and {Roegiers}, T. and {Rybizki}, J. and {Sarro}, L.~M. and {Siopis}, C. and {Smith}, M. and {Sozzetti}, A. and {Ulla}, A. and {Utrilla}, E. and {van Leeuwen}, M. and {van Reeven}, W. and {Abbas}, U. and {Abreu Aramburu}, A. and {Accart}, S. and {Aerts}, C. and {Aguado}, J.~J. and {Ajaj}, M. and {Altavilla}, G. and {{\'A}lvarez}, M.~A. and {{\'A}lvarez Cid-Fuentes}, J. and {Alves}, J. and {Anderson}, R.~I. and {Anglada Varela}, E. and {Antoja}, T. and {Audard}, M. and {Baines}, D. and {Baker}, S.~G. and {Balaguer-N{\'u}{\~n}ez}, L. and {Balbinot}, E. and {Balog}, Z. and {Barache}, C. and {Barbato}, D. and {Barros}, M. and {Barstow}, M.~A. and {Bartolom{\'e}}, S. and {Bassilana}, J. -L. and {Bauchet}, N. and {Baudesson-Stella}, A. and {Becciani}, U. and {Bellazzini}, M. and {Bernet}, M. and {Bertone}, S. and {Bianchi}, L. and {Blanco-Cuaresma}, S. and {Boch}, T. and {Bombrun}, A. and {Bossini}, D. and {Bouquillon}, S. and {Bragaglia}, A. and {Bramante}, L. and {Breedt}, E. and {Bressan}, A. and {Brouillet}, N. and {Bucciarelli}, B. and {Burlacu}, A. and {Busonero}, D. and {Butkevich}, A.~G. and {Buzzi}, R. and {Caffau}, E. and {Cancelliere}, R. and {C{\'a}novas}, H. and {Cantat-Gaudin}, T. and {Carballo}, R. and {Carlucci}, T. and {Carnerero}, M.~I. and {Carrasco}, J.~M. and {Casamiquela}, L. and {Castellani}, M. and {Castro-Ginard}, A. and {Castro Sampol}, P. and {Chaoul}, L. and {Charlot}, P. and {Chemin}, L. and {Chiavassa}, A. and {Cioni}, M. -R.~L. and {Comoretto}, G. and {Cooper}, W.~J. and {Cornez}, T. and {Cowell}, S. and {Crifo}, F. and {Crosta}, M. and {Crowley}, C. and {Dafonte}, C. and {Dapergolas}, A. and {David}, M. and {David}, P.},
        title = "{Gaia Early Data Release 3. Summary of the contents and survey properties}",
      journal = {\aap},
         year = 2021,
        month = may,
       volume = {649},
          eid = {A1},
        pages = {A1},
          doi = {10.1051/0004-6361/202039657},
archivePrefix = {arXiv},
       eprint = {2012.01533},
 primaryClass = {astro-ph.GA},
       adsurl = {https://ui.adsabs.harvard.edu/abs/2021A&A...649A...1G}
}

@ARTICLE{2021AJ....161..147B,
       author = {{Bailer-Jones}, C.~A.~L. and {Rybizki}, J. and {Fouesneau}, M. and {Demleitner}, M. and {Andrae}, R.},
        title = "{Estimating Distances from Parallaxes. V. Geometric and Photogeometric Distances to 1.47 Billion Stars in Gaia Early Data Release 3}",
      journal = {\aj},
         year = 2021,
        month = mar,
       volume = {161},
       number = {3},
          eid = {147},
        pages = {147},
          doi = {10.3847/1538-3881/abd806},
archivePrefix = {arXiv},
       eprint = {2012.05220},
 primaryClass = {astro-ph.SR},
       adsurl = {https://ui.adsabs.harvard.edu/abs/2021AJ....161..147B}
}

@ARTICLE{2019ApJ...887...93G,
       author = {{Green}, Gregory M. and {Schlafly}, Edward and {Zucker}, Catherine and {Speagle}, Joshua S. and {Finkbeiner}, Douglas},
        title = "{A 3D Dust Map Based on Gaia, Pan-STARRS 1, and 2MASS}",
      journal = {\apj},
         year = 2019,
        month = dec,
       volume = {887},
       number = {1},
          eid = {93},
        pages = {93},
          doi = {10.3847/1538-4357/ab5362},
archivePrefix = {arXiv},
       eprint = {1905.02734},
 primaryClass = {astro-ph.GA},
       adsurl = {https://ui.adsabs.harvard.edu/abs/2019ApJ...887...93G}
}

@ARTICLE{1998ApJ...500..525S,
       author = {{Schlegel}, David J. and {Finkbeiner}, Douglas P. and {Davis}, Marc},
        title = "{Maps of Dust Infrared Emission for Use in Estimation of Reddening and Cosmic Microwave Background Radiation Foregrounds}",
      journal = {\apj},
         year = 1998,
        month = jun,
       volume = {500},
       number = {2},
        pages = {525-553},
          doi = {10.1086/305772},
archivePrefix = {arXiv},
       eprint = {astro-ph/9710327},
 primaryClass = {astro-ph},
       adsurl = {https://ui.adsabs.harvard.edu/abs/1998ApJ...500..525S}
}

@ARTICLE{1999PASP..111...63F,
       author = {{Fitzpatrick}, Edward L.},
        title = "{Correcting for the Effects of Interstellar Extinction}",
      journal = {\pasp},
         year = 1999,
        month = jan,
       volume = {111},
       number = {755},
        pages = {63-75},
          doi = {10.1086/316293},
archivePrefix = {arXiv},
       eprint = {astro-ph/9809387},
 primaryClass = {astro-ph},
       adsurl = {https://ui.adsabs.harvard.edu/abs/1999PASP..111...63F}
}

@ARTICLE{2019ApJS..244....8Z,
       author = {{Zhong}, Jing and {Li}, Jing and {Carlin}, Jeffrey L. and {Chen}, Li and {Mendez}, Rene A. and {Hou}, Jinliang},
        title = "{Value-added Catalogs of M-type Stars in LAMOST DR5}",
      journal = {\apjs},
         year = 2019,
        month = sep,
       volume = {244},
       number = {1},
          eid = {8},
        pages = {8},
          doi = {10.3847/1538-4365/ab3859},
archivePrefix = {arXiv},
       eprint = {1908.01128},
 primaryClass = {astro-ph.GA},
       adsurl = {https://ui.adsabs.harvard.edu/abs/2019ApJS..244....8Z}
}

@ARTICLE{2025ApJS..277...47Z,
       author = {{Zhang}, Shuo and {Zhang}, Hua-Wei and {Ting}, Yuan-Sen and {Wang}, Rui and {O'Briain}, Teaghan and {Jones}, Hugh R.~A. and {Homeier}, Derek and {Luo}, A. -Li},
        title = "{Half a Million M Dwarf Stars Characterized Using Domain-adapted Spectral Analysis}",
      journal = {\apjs},
         year = 2025,
        month = apr,
       volume = {277},
       number = {2},
          eid = {47},
        pages = {47},
          doi = {10.3847/1538-4365/adb614},
archivePrefix = {arXiv},
       eprint = {2502.01910},
 primaryClass = {astro-ph.SR},
       adsurl = {https://ui.adsabs.harvard.edu/abs/2025ApJS..277...47Z}
}

@ARTICLE{2024ApJ...966...69L,
       author = {{Li}, Xue and {Wang}, Song and {Han}, Henggeng and {Yang}, Huiqin and {Zheng}, Chuanjie and {Huang}, Yang and {Liu}, Jifeng},
        title = "{Ultraviolet and Chromospheric Activity and Habitability of M Stars}",
      journal = {\apj},
         year = 2024,
        month = may,
       volume = {966},
       number = {1},
          eid = {69},
        pages = {69},
          doi = {10.3847/1538-4357/ad3038},
archivePrefix = {arXiv},
       eprint = {2402.17384},
 primaryClass = {astro-ph.SR},
       adsurl = {https://ui.adsabs.harvard.edu/abs/2024ApJ...966...69L}
}

@ARTICLE{1978PASP...90..267V,
       author = {{Vaughan}, A.~H. and {Preston}, G.~W. and {Wilson}, O.~C.},
        title = "{Flux measurements of Ca II and K emission.}",
      journal = {\pasp},
         year = 1978,
        month = jun,
       volume = {90},
        pages = {267-274},
          doi = {10.1086/130324},
       adsurl = {https://ui.adsabs.harvard.edu/abs/1978PASP...90..267V}
}

@ARTICLE{2025ApJ...984....2H,
       author = {{Han}, Henggeng and {Wang}, Song and {Li}, Xue and {Zheng}, Chuanjie and {Liu}, Jifeng},
        title = "{Impact of Spectral Resolution on S-index and Its Application to Spectroscopic Surveys}",
      journal = {\apj},
         year = 2025,
        month = may,
       volume = {984},
       number = {1},
          eid = {2},
        pages = {2},
          doi = {10.3847/1538-4357/adc600},
archivePrefix = {arXiv},
       eprint = {2503.20165},
 primaryClass = {astro-ph.SR},
       adsurl = {https://ui.adsabs.harvard.edu/abs/2025ApJ...984....2H}
}

@ARTICLE{2018A&A...616A.108B,
       author = {{Boro Saikia}, S. and {Marvin}, C.~J. and {Jeffers}, S.~V. and {Reiners}, A. and {Cameron}, R. and {Marsden}, S.~C. and {Petit}, P. and {Warnecke}, J. and {Yadav}, A.~P.},
        title = "{Chromospheric activity catalogue of 4454 cool stars. Questioning the active branch of stellar activity cycles}",
      journal = {\aap},
         year = 2018,
        month = aug,
       volume = {616},
          eid = {A108},
        pages = {A108},
          doi = {10.1051/0004-6361/201629518},
archivePrefix = {arXiv},
       eprint = {1803.11123},
 primaryClass = {astro-ph.SR},
       adsurl = {https://ui.adsabs.harvard.edu/abs/2018A&A...616A.108B}
}

@ARTICLE{1984ApJ...279..763N,
       author = {{Noyes}, R.~W. and {Hartmann}, L.~W. and {Baliunas}, S.~L. and {Duncan}, D.~K. and {Vaughan}, A.~H.},
        title = "{Rotation, convection, and magnetic activity in lower main-sequence stars.}",
      journal = {\apj},
         year = 1984,
        month = apr,
       volume = {279},
        pages = {763-777},
          doi = {10.1086/161945},
       adsurl = {https://ui.adsabs.harvard.edu/abs/1984ApJ...279..763N}
}

@ARTICLE{1982A&A...107...31M,
       author = {{Middelkoop}, F.},
        title = "{Magnetic structure in cool stars. IV - Rotation and CA II H and K emission of main-sequence stars}",
      journal = {\aap},
         year = 1982,
        month = mar,
       volume = {107},
       number = {1},
        pages = {31-35},
       adsurl = {https://ui.adsabs.harvard.edu/abs/1982A&A...107...31M}
}

@ARTICLE{2013ApJS..208....9P,
       author = {{Pecaut}, Mark J. and {Mamajek}, Eric E.},
        title = "{Intrinsic Colors, Temperatures, and Bolometric Corrections of Pre-main-sequence Stars}",
      journal = {\apjs},
         year = 2013,
        month = sep,
       volume = {208},
       number = {1},
          eid = {9},
        pages = {9},
          doi = {10.1088/0067-0049/208/1/9},
archivePrefix = {arXiv},
       eprint = {1307.2657},
 primaryClass = {astro-ph.SR},
       adsurl = {https://ui.adsabs.harvard.edu/abs/2013ApJS..208....9P}
}

@ARTICLE{2013MNRAS.432.1203M,
       author = {{McQuillan}, A. and {Aigrain}, S. and {Mazeh}, T.},
        title = "{Measuring the rotation period distribution of field M dwarfs with Kepler}",
      journal = {\mnras},
         year = 2013,
        month = jun,
       volume = {432},
       number = {2},
        pages = {1203-1216},
          doi = {10.1093/mnras/stt536},
archivePrefix = {arXiv},
       eprint = {1303.6787},
 primaryClass = {astro-ph.SR},
       adsurl = {https://ui.adsabs.harvard.edu/abs/2013MNRAS.432.1203M}
}

@ARTICLE{2014ApJS..211...24M,
       author = {{McQuillan}, A. and {Mazeh}, T. and {Aigrain}, S.},
        title = "{Rotation Periods of 34,030 Kepler Main-sequence Stars: The Full Autocorrelation Sample}",
      journal = {\apjs},
         year = 2014,
        month = apr,
       volume = {211},
       number = {2},
          eid = {24},
        pages = {24},
          doi = {10.1088/0067-0049/211/2/24},
archivePrefix = {arXiv},
       eprint = {1402.5694},
 primaryClass = {astro-ph.SR},
       adsurl = {https://ui.adsabs.harvard.edu/abs/2014ApJS..211...24M}
}

@ARTICLE{2019ApJS..244...21S,
       author = {{Santos}, A.~R.~G. and {Garc{\'\i}a}, R.~A. and {Mathur}, S. and {Bugnet}, L. and {van Saders}, J.~L. and {Metcalfe}, T.~S. and {Simonian}, G.~V.~A. and {Pinsonneault}, M.~H.},
        title = "{Surface Rotation and Photometric Activity for Kepler Targets. I. M and K Main-sequence Stars}",
      journal = {\apjs},
         year = 2019,
        month = sep,
       volume = {244},
       number = {1},
          eid = {21},
        pages = {21},
          doi = {10.3847/1538-4365/ab3b56},
archivePrefix = {arXiv},
       eprint = {1908.05222},
 primaryClass = {astro-ph.SR},
       adsurl = {https://ui.adsabs.harvard.edu/abs/2019ApJS..244...21S}
}

@ARTICLE{2020A&A...635A..43R,
       author = {{Reinhold}, Timo and {Hekker}, Saskia},
        title = "{Stellar rotation periods from K2 Campaigns 0-18. Evidence for rotation period bimodality and simultaneous variability decrease}",
      journal = {\aap},
         year = 2020,
        month = mar,
       volume = {635},
          eid = {A43},
        pages = {A43},
          doi = {10.1051/0004-6361/201936887},
archivePrefix = {arXiv},
       eprint = {2001.08214},
 primaryClass = {astro-ph.SR},
       adsurl = {https://ui.adsabs.harvard.edu/abs/2020A&A...635A..43R}
}

@ARTICLE{2025ApJS..276...57G,
       author = {{Gao}, Xinyi and {Chen}, Xiaodian and {Wang}, Shu and {Liu}, Jifeng},
        title = "{Classification of Periodic Variable Stars from TESS}",
      journal = {\apjs},
         year = 2025,
        month = feb,
       volume = {276},
       number = {2},
          eid = {57},
        pages = {57},
          doi = {10.3847/1538-4365/ad9dd6},
archivePrefix = {arXiv},
       eprint = {2412.06175},
 primaryClass = {astro-ph.SR},
       adsurl = {https://ui.adsabs.harvard.edu/abs/2025ApJS..276...57G}
}

@ARTICLE{1955ApJ...122..293P,
       author = {{Parker}, Eugene N.},
        title = "{Hydromagnetic Dynamo Models.}",
      journal = {\apj},
         year = 1955,
        month = sep,
       volume = {122},
        pages = {293},
          doi = {10.1086/146087},
       adsurl = {https://ui.adsabs.harvard.edu/abs/1955ApJ...122..293P}
}

@ARTICLE{2017ApJ...838..161S,
       author = {{Spada}, F. and {Demarque}, P. and {Kim}, Y. -C. and {Boyajian}, T.~S. and {Brewer}, J.~M.},
        title = "{The Yale-Potsdam Stellar Isochrones}",
      journal = {\apj},
         year = 2017,
        month = apr,
       volume = {838},
       number = {2},
          eid = {161},
        pages = {161},
          doi = {10.3847/1538-4357/aa661d},
archivePrefix = {arXiv},
       eprint = {1703.03975},
 primaryClass = {astro-ph.SR},
       adsurl = {https://ui.adsabs.harvard.edu/abs/2017ApJ...838..161S}
}

@ARTICLE{2009ApJ...692..538R,
       author = {{Reiners}, A. and {Basri}, G. and {Browning}, M.},
        title = "{Evidence for Magnetic Flux Saturation in Rapidly Rotating M Stars}",
      journal = {\apj},
         year = 2009,
        month = feb,
       volume = {692},
       number = {1},
        pages = {538-545},
          doi = {10.1088/0004-637X/692/1/538},
archivePrefix = {arXiv},
       eprint = {0810.5139},
 primaryClass = {astro-ph},
       adsurl = {https://ui.adsabs.harvard.edu/abs/2009ApJ...692..538R}
}

@ARTICLE{2021A&A...649A..96J,
       author = {{Johnstone}, C.~P. and {Bartel}, M. and {G{\"u}del}, M.},
        title = "{The active lives of stars: A complete description of the rotation and XUV evolution of F, G, K, and M dwarfs}",
      journal = {\aap},
         year = 2021,
        month = may,
       volume = {649},
          eid = {A96},
        pages = {A96},
          doi = {10.1051/0004-6361/202038407},
archivePrefix = {arXiv},
       eprint = {2009.07695},
 primaryClass = {astro-ph.SR},
       adsurl = {https://ui.adsabs.harvard.edu/abs/2021A&A...649A..96J}
}

@ARTICLE{2025MNRAS.539.1922S,
       author = {{Stuart}, Kieran A. and {Gregory}, Scott G.},
        title = "{Modelling the emergence and evolution of the rotation{\textendash}activity relation}",
      journal = {\mnras},
         year = 2025,
        month = may,
       volume = {539},
       number = {3},
        pages = {1922-1943},
          doi = {10.1093/mnras/staf589},
archivePrefix = {arXiv},
       eprint = {2504.07263},
 primaryClass = {astro-ph.SR},
       adsurl = {https://ui.adsabs.harvard.edu/abs/2025MNRAS.539.1922S}
}

@ARTICLE{2023ApJS..264...12H,
       author = {{Han}, Henggeng and {Wang}, Song and {Bai}, Yu and {Yang}, Huiqin and {Fang}, Xiangsong and {Liu}, Jifeng},
        title = "{Stellar Chromospheric Activities Revealed from the LAMOST-K2 Time-domain Survey}",
      journal = {\apjs},
         year = 2023,
        month = jan,
       volume = {264},
       number = {1},
          eid = {12},
        pages = {12},
          doi = {10.3847/1538-4365/ac9eac},
archivePrefix = {arXiv},
       eprint = {2210.16830},
 primaryClass = {astro-ph.SR},
       adsurl = {https://ui.adsabs.harvard.edu/abs/2023ApJS..264...12H}
}

@ARTICLE{2021ApJ...910..110L,
       author = {{Lehtinen}, Jyri J. and {K{\"a}pyl{\"a}}, Maarit J. and {Olspert}, Nigul and {Spada}, Federico},
        title = "{A Knee Point in the Rotation-Activity Scaling of Late-type Stars with a Connection to Dynamo Transitions}",
      journal = {\apj},
         year = 2021,
        month = apr,
       volume = {910},
       number = {2},
          eid = {110},
        pages = {110},
          doi = {10.3847/1538-4357/abe621},
archivePrefix = {arXiv},
       eprint = {2007.00040},
 primaryClass = {astro-ph.SR},
       adsurl = {https://ui.adsabs.harvard.edu/abs/2021ApJ...910..110L}
}

@ARTICLE{2024A&A...684A.121F,
       author = {{Freund}, S. and {Czesla}, S. and {Predehl}, P. and {Robrade}, J. and {Salvato}, M. and {Schneider}, P.~C. and {Starck}, H. and {Wolf}, J. and {Schmitt}, J.~H.~M.~M.},
        title = "{The SRG/eROSITA all-sky survey. Identifying the coronal content with HamStar}",
      journal = {\aap},
         year = 2024,
        month = apr,
       volume = {684},
          eid = {A121},
        pages = {A121},
          doi = {10.1051/0004-6361/202348278},
archivePrefix = {arXiv},
       eprint = {2401.17282},
 primaryClass = {astro-ph.SR},
       adsurl = {https://ui.adsabs.harvard.edu/abs/2024A&A...684A.121F}
}

@ARTICLE{2010ApJ...721..675B,
       author = {{Barnes}, Sydney A. and {Kim}, Yong-Cheol},
        title = "{Angular Momentum Loss from Cool Stars: An Empirical Expression and Connection to Stellar Activity}",
      journal = {\apj},
         year = 2010,
        month = sep,
       volume = {721},
       number = {1},
        pages = {675-685},
          doi = {10.1088/0004-637X/721/1/675},
archivePrefix = {arXiv},
       eprint = {1104.2350},
 primaryClass = {astro-ph.SR},
       adsurl = {https://ui.adsabs.harvard.edu/abs/2010ApJ...721..675B}
}

@ARTICLE{2018ApJ...862...33A,
       author = {{Ag{\"u}eros}, M.~A. and {Bowsher}, E.~C. and {Bochanski}, J.~J. and {Cargile}, P.~A. and {Covey}, K.~R. and {Douglas}, S.~T. and {Kraus}, A. and {Kundert}, A. and {Law}, N.~M. and {Ahmadi}, A. and {Arce}, H.~G.},
        title = "{A New Look at an Old Cluster: The Membership, Rotation, and Magnetic Activity of Low-mass Stars in the 1.3 Gyr Old Open Cluster NGC 752}",
      journal = {\apj},
         year = 2018,
        month = jul,
       volume = {862},
       number = {1},
          eid = {33},
        pages = {33},
          doi = {10.3847/1538-4357/aac6ed},
archivePrefix = {arXiv},
       eprint = {1804.02016},
 primaryClass = {astro-ph.SR},
       adsurl = {https://ui.adsabs.harvard.edu/abs/2018ApJ...862...33A}
}

@ARTICLE{2007ApJ...669.1167B,
       author = {{Barnes}, Sydney A.},
        title = "{Ages for Illustrative Field Stars Using Gyrochronology: Viability, Limitations, and Errors}",
      journal = {\apj},
         year = 2007,
        month = nov,
       volume = {669},
       number = {2},
        pages = {1167-1189},
          doi = {10.1086/519295},
archivePrefix = {arXiv},
       eprint = {0704.3068},
 primaryClass = {astro-ph},
       adsurl = {https://ui.adsabs.harvard.edu/abs/2007ApJ...669.1167B}
}

@ARTICLE{2010ApJ...722..222B,
       author = {{Barnes}, Sydney A.},
        title = "{A Simple Nonlinear Model for the Rotation of Main-sequence Cool Stars. I. Introduction, Implications for Gyrochronology, and Color-Period Diagrams}",
      journal = {\apj},
         year = 2010,
        month = oct,
       volume = {722},
       number = {1},
        pages = {222-234},
          doi = {10.1088/0004-637X/722/1/222},
       adsurl = {https://ui.adsabs.harvard.edu/abs/2010ApJ...722..222B}
}

@ARTICLE{2022ApJ...938..118D,
       author = {{Dungee}, Ryan and {van Saders}, Jennifer and {Gaidos}, Eric and {Chun}, Mark and {Garc{\'\i}a}, Rafael A. and {Magnier}, Eugene A. and {Mathur}, Savita and {Santos}, {\^A}ngela R.~G.},
        title = "{A 4 Gyr M-dwarf Gyrochrone from CFHT/MegaPrime Monitoring of the Open Cluster M67}",
      journal = {\apj},
         year = 2022,
        month = oct,
       volume = {938},
       number = {2},
          eid = {118},
        pages = {118},
          doi = {10.3847/1538-4357/ac90be},
archivePrefix = {arXiv},
       eprint = {2211.01377},
 primaryClass = {astro-ph.SR},
       adsurl = {https://ui.adsabs.harvard.edu/abs/2022ApJ...938..118D}
}

@ARTICLE{2021AJ....161..189L,
       author = {{Lu}, Yuxi Lucy and {Angus}, Ruth and {Curtis}, Jason L. and {David}, Trevor J. and {Kiman}, Rocio},
        title = "{Gyro-kinematic Ages for around 30,000 Kepler Stars}",
      journal = {\aj},
         year = 2021,
        month = apr,
       volume = {161},
       number = {4},
          eid = {189},
        pages = {189},
          doi = {10.3847/1538-3881/abe4d6},
archivePrefix = {arXiv},
       eprint = {2102.01772},
 primaryClass = {astro-ph.SR},
       adsurl = {https://ui.adsabs.harvard.edu/abs/2021AJ....161..189L}
}

@ARTICLE{2013PASP..125..306F,
       author = {{Foreman-Mackey}, Daniel and {Hogg}, David W. and {Lang}, Dustin and {Goodman}, Jonathan},
        title = "{emcee: The MCMC Hammer}",
      journal = {\pasp},
         year = 2013,
        month = mar,
       volume = {125},
       number = {925},
        pages = {306},
          doi = {10.1086/670067},
archivePrefix = {arXiv},
       eprint = {1202.3665},
 primaryClass = {astro-ph.IM},
       adsurl = {https://ui.adsabs.harvard.edu/abs/2013PASP..125..306F}
}

@ARTICLE{2023AJ....166...14B,
       author = {{Boyle}, Andrew W. and {Bouma}, Luke G.},
        title = "{Stellar Rotation and Structure of the {\ensuremath{\alpha}} Persei Complex: When Does Gyrochronology Start to Work?}",
      journal = {\aj},
         year = 2023,
        month = jul,
       volume = {166},
       number = {1},
          eid = {14},
        pages = {14},
          doi = {10.3847/1538-3881/acd3e8},
archivePrefix = {arXiv},
       eprint = {2211.09822},
 primaryClass = {astro-ph.SR},
       adsurl = {https://ui.adsabs.harvard.edu/abs/2023AJ....166...14B}
}

@dataset{https://doi.org/10.17909/t9-nmc8-f686,
doi = {10.17909/T9-NMC8-F686},
url = {http://archive.stsci.edu/doi/resolve/resolve.html?doi=10.17909/t9-nmc8-f686},
author = {{TESS Team}},
title = {TESS Light Curves - All Sectors},
publisher = {STScI/MAST},
year = {2021}
}

@dataset{https://doi.org/10.17909/t9488n,
doi = {10.17909/T9488N},
url = {http://archive.stsci.edu/doi/resolve/resolve.html?doi=10.17909/T9488N},
author = {{STScI}},
title = {Kepler LC, Q0-Q17},
publisher = {STScI/MAST},
year = {2016}
}

@dataset{https://doi.org/10.17909/t9ws3r,
doi = {10.17909/T9WS3R},
url = {http://archive.stsci.edu/doi/resolve/resolve.html?doi=10.17909/T9WS3R},
author = {{STScI}},
title = {K2 Light Curves (all)},
publisher = {STScI/MAST},
year = {2016}
}
\bibliographystyle{aasjournalv7}

\clearpage
\begin{appendix}

\section{Comparisons of Activity Indices and rotation--activity relation}
\renewcommand\thefigure{\Alph{section}\arabic{figure}}
\renewcommand\thetable{\Alph{section}\arabic{table}}
\setcounter{figure}{0}
\setcounter{table}{0}  

\begin{figure*}[h]
\centering
\includegraphics[width=0.8\textwidth]{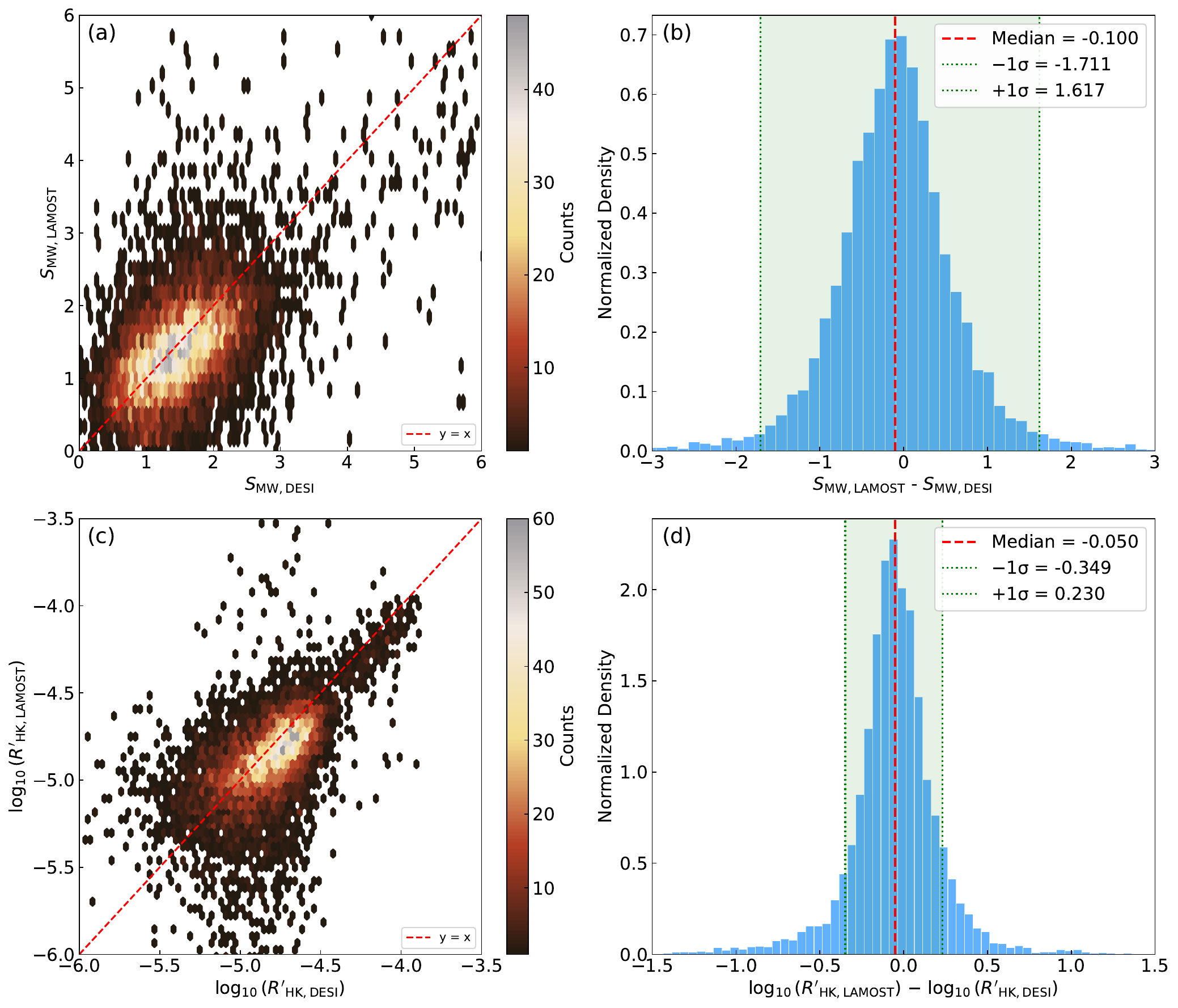}
\caption{Panel (a): Comparison between the calibrated $S_{\rm{MW}}$ from LAMOST and DESI. Panel (b): Histogram of differences between the calibrated $S_{\rm{MW}}$. Panel (c): Comparison between the $R_{\rm{HK}}^{'}$ indices from LAMOST and DESI. Panel (d): Histogram of differences between $R_{\rm{HK}}^{'}$ indices from LAMOST and DESI.}
\label{scomp.fig}
\end{figure*}

In Figure \ref{scomp.fig}, we give the comparisons of LAMOST and DESI activity indices, which show good agreement with each other.

\begin{figure*}[h]
\centering
\includegraphics[width=0.8\textwidth]{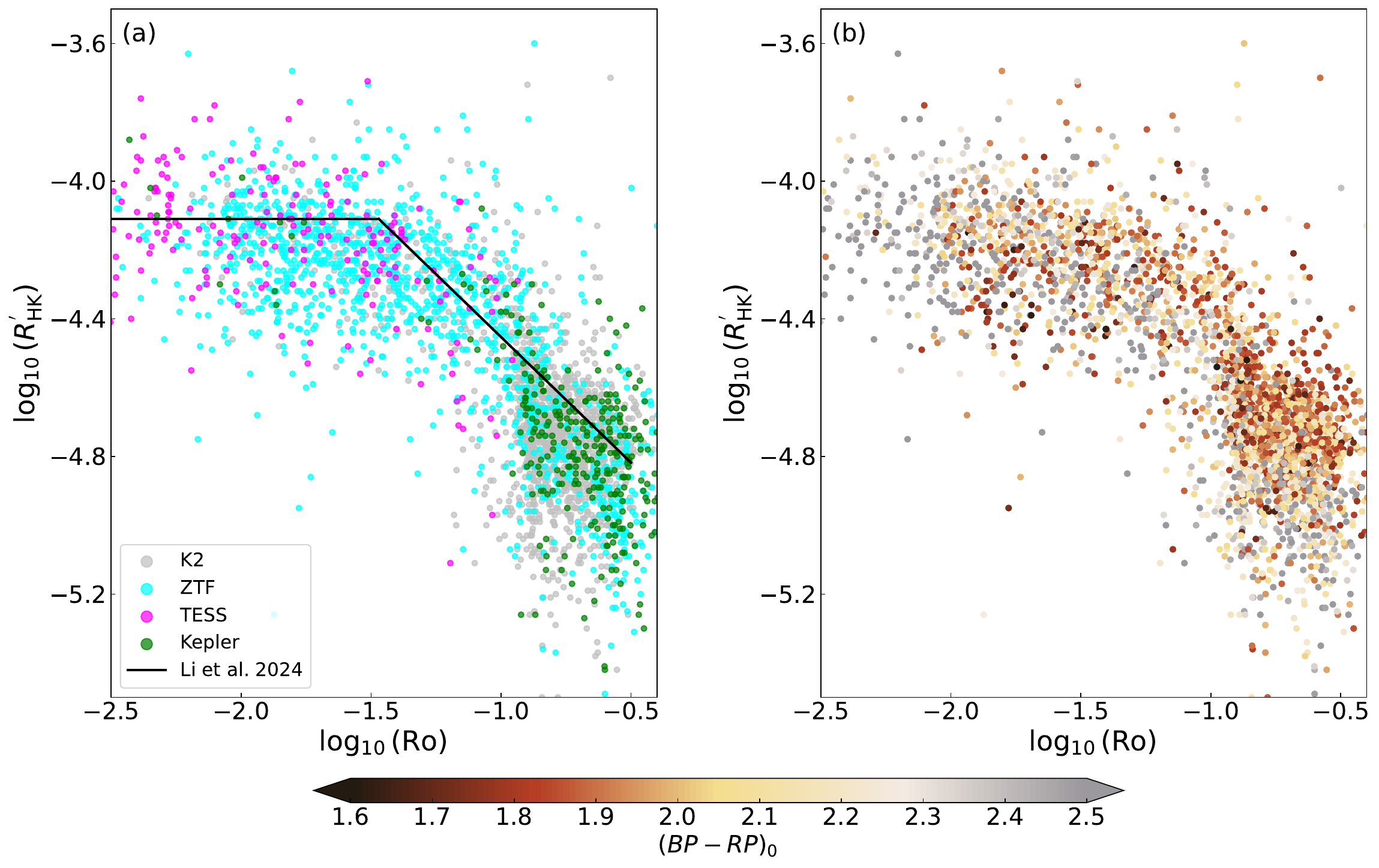}
\caption{$R_{\rm{HK}}^{'}$--Ro relation of M dwarfs. Panel (a): Different colors represent different sky surveys. Black line represents best-fit model from \cite{2024ApJ...966...69L}. Panel (b): Same as panel (a) with different colors represent different $(BP-RP)_{0}$. }
\label{ar_all.fig}
\end{figure*}

\clearpage
\section{Posterior probability distributions of the fitting results}
\renewcommand\thefigure{\Alph{section}\arabic{figure}}
\renewcommand\thetable{\Alph{section}\arabic{table}}
\setcounter{figure}{0}
\setcounter{table}{0}

In this work, we applied uniform prior to the fitting parameters. The likelihood is written as:
\begin{equation}
\label{like_correct}
    \ln p(\boldsymbol{y}|\boldsymbol{x},\boldsymbol{\sigma},s,\boldsymbol{\theta}) = -\frac{1}{2} \sum_{i=1}^{N} \left[ \frac{[y_i - f(x_i, \boldsymbol{\theta})]^{2}}{\sigma_i^{2}+s^{2}} + \ln(\sigma_i^{2}+s^{2}) \right].
\end{equation}
Here $f(x, \boldsymbol{\theta})$ is the four-segment piecewise model, $\boldsymbol{\theta} = (x_0, x_1, x_2, a_0, a_1, a_2, b)$. s is the intrinsic scatter of the rotation--activity relation. The best-fit parameters are calculated using the maximum likelihood estimation. The likelihood function for the binned median approach follows the same formulation, with the exception that it omits the intrinsic scatter.

\begin{figure*}[h]
\centering

\includegraphics[width=0.85\textwidth]{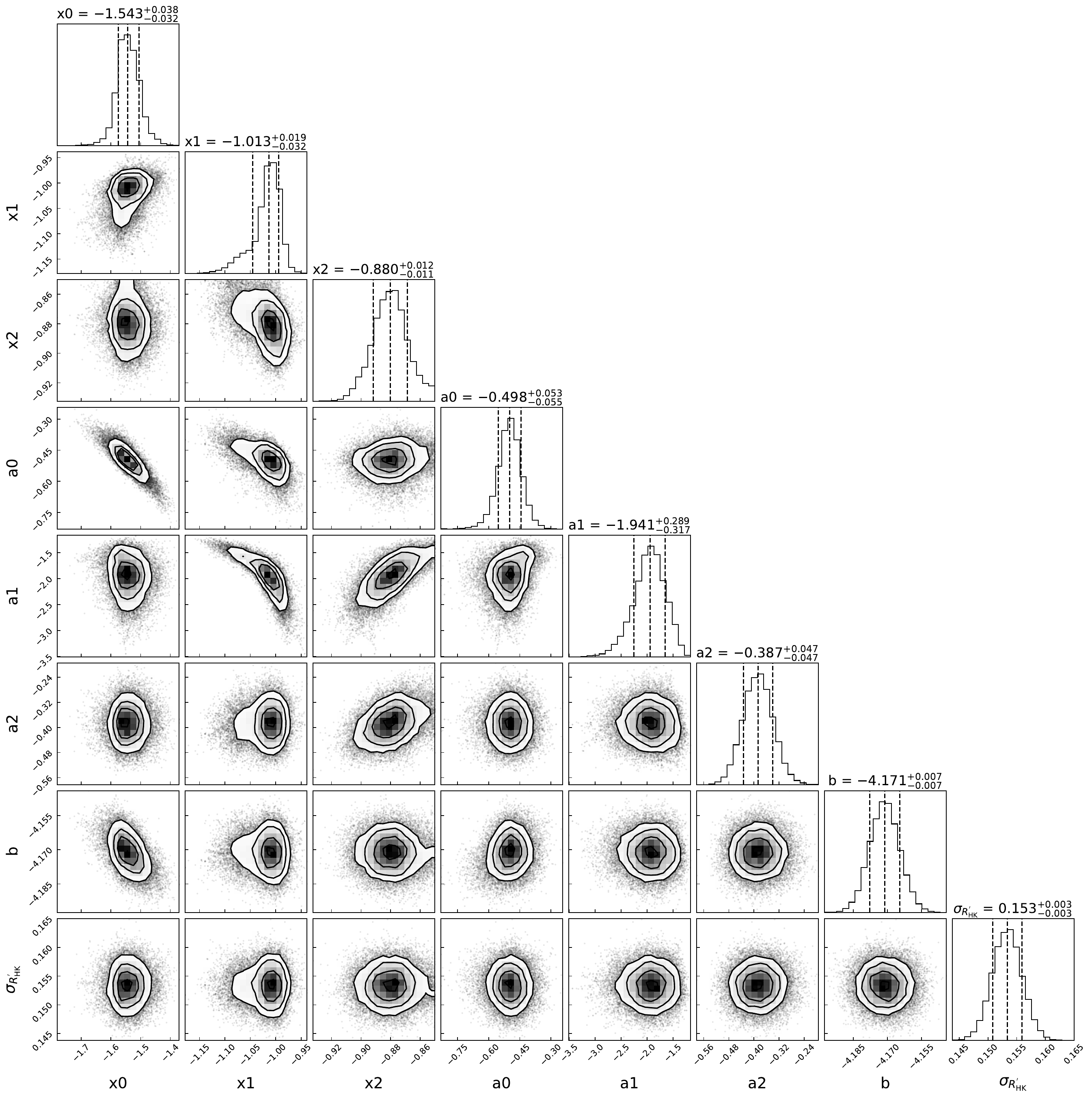}

\caption{Posterior probability distributions of the fitting results corresponding to all points fitting method.}
\label{post.fig}
\end{figure*}

\begin{figure*}[h]
\centering

\includegraphics[width=0.85\textwidth]{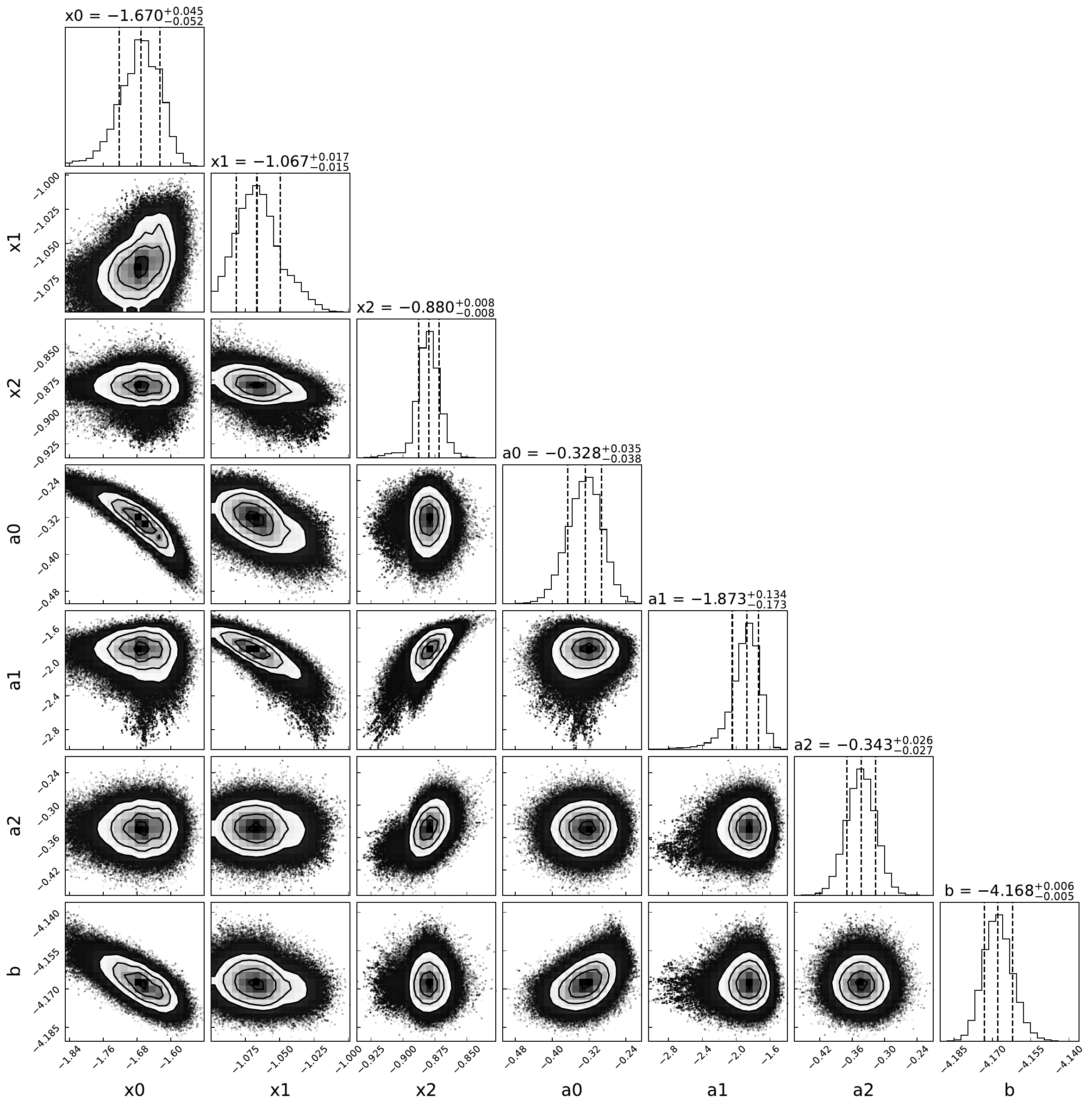}

\caption{Posterior probability distributions of the fitting results corresponding to binned median fitting method.}
\label{post_binned.fig}
\end{figure*}

\clearpage
\section{Fitting Results of rotation--activity relation in linear-log scale}
\renewcommand\thefigure{\Alph{section}\arabic{figure}}
\renewcommand\thetable{\Alph{section}\arabic{table}}
\setcounter{figure}{0}
\setcounter{table}{0}

\begin{figure*}[h]
\centering
\subfigure[]{
\includegraphics[width=0.85\textwidth]{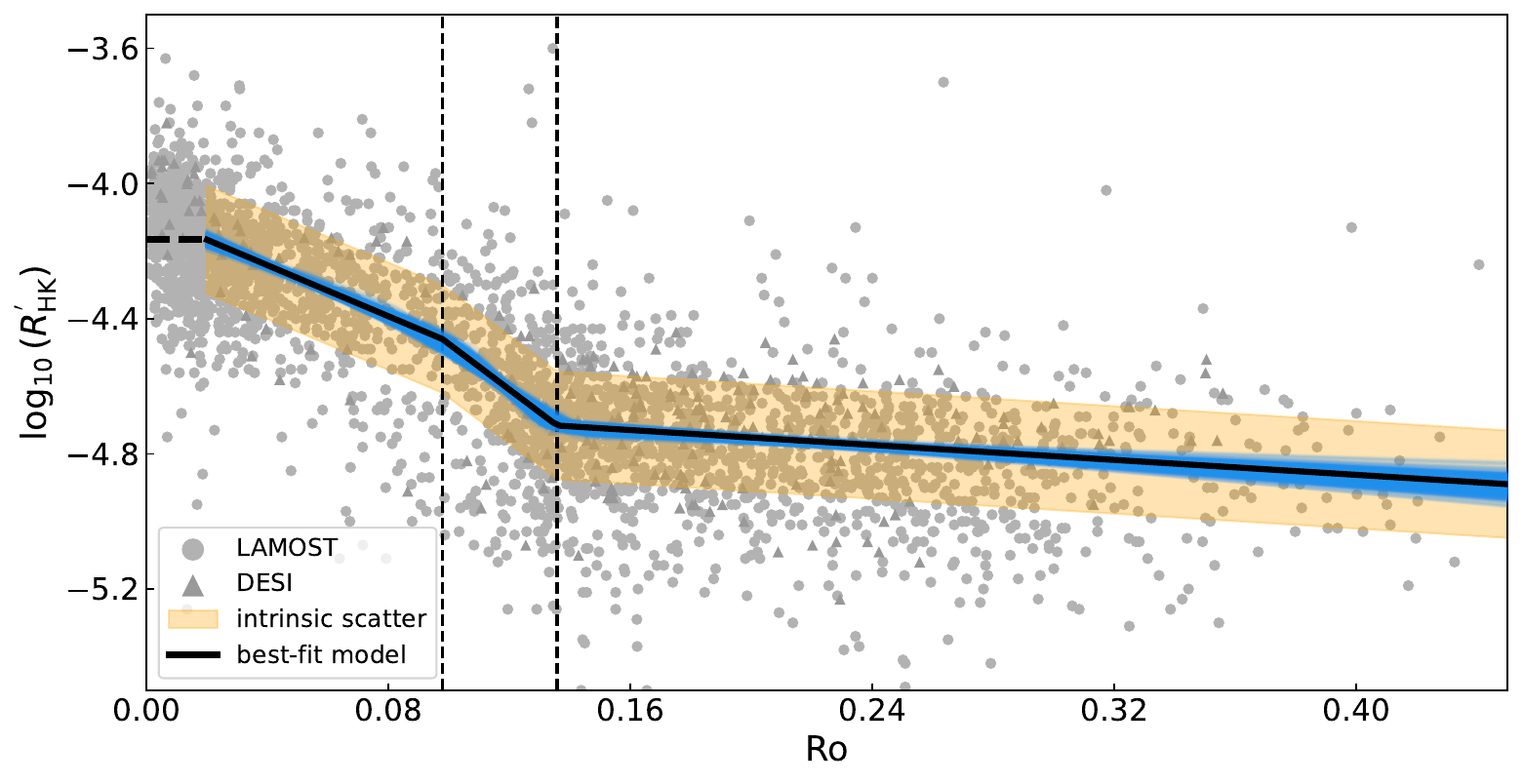}}
\subfigure[]{
\includegraphics[width=0.85\textwidth]{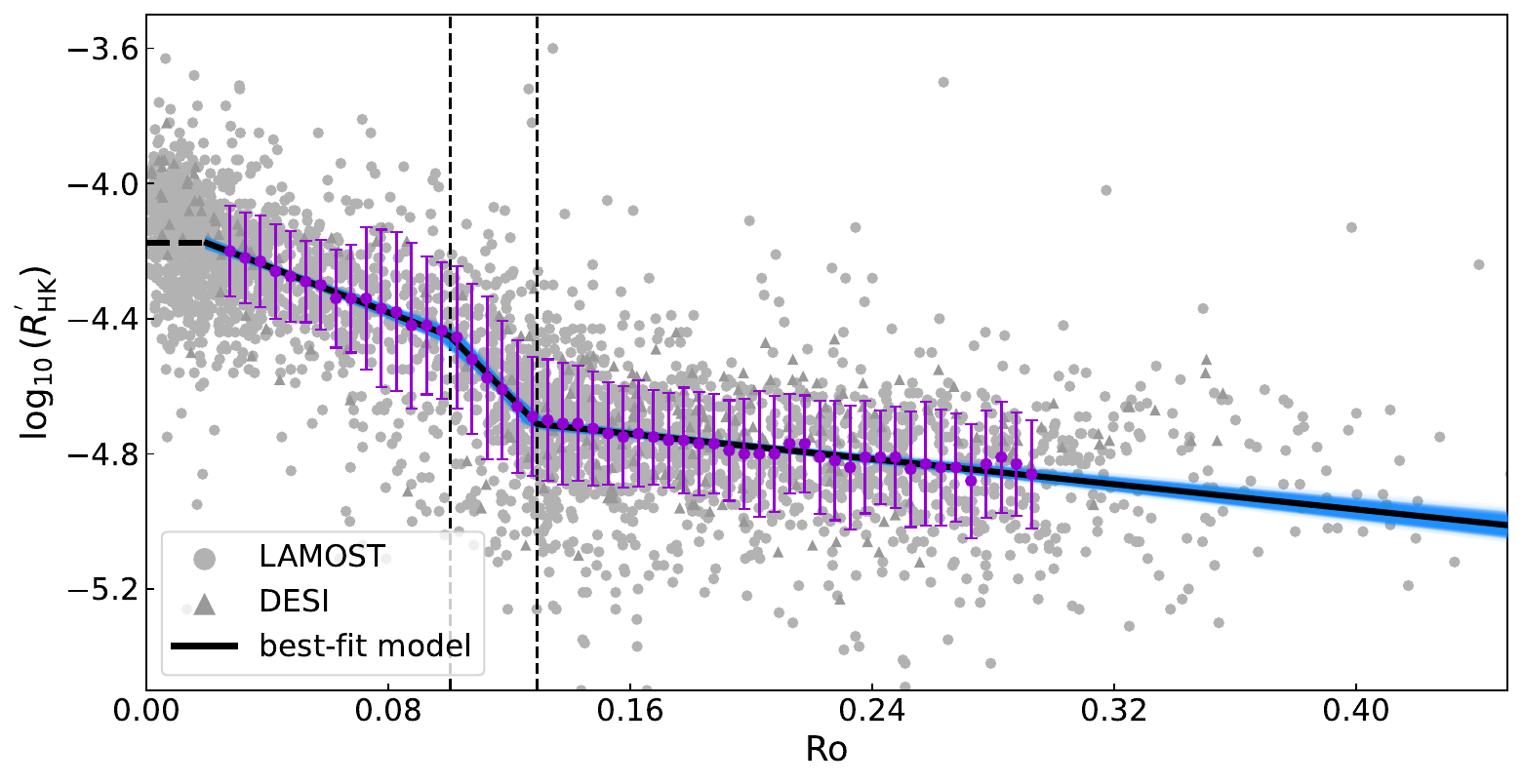}}
\caption{Panel (a): $R_{\rm{HK}}^{'}$--Ro relation of M dwarfs in linear-log scale. Black line is the best-fit model. Black vertical dashed lines mark the two knee points. Blue lines are 1000 models randomly extracted from posterior probability distributions. Shaded area represents intrinsic scatter of the log$_{10}(R_{\rm{HK}}^{'})$. Panel (b): Same as panel (a) but for binned data fitting method.}
\label{ar_fit_piece_linear.fig}
\end{figure*}

\begin{figure*}[h]
\centering

\includegraphics[width=0.85\textwidth]{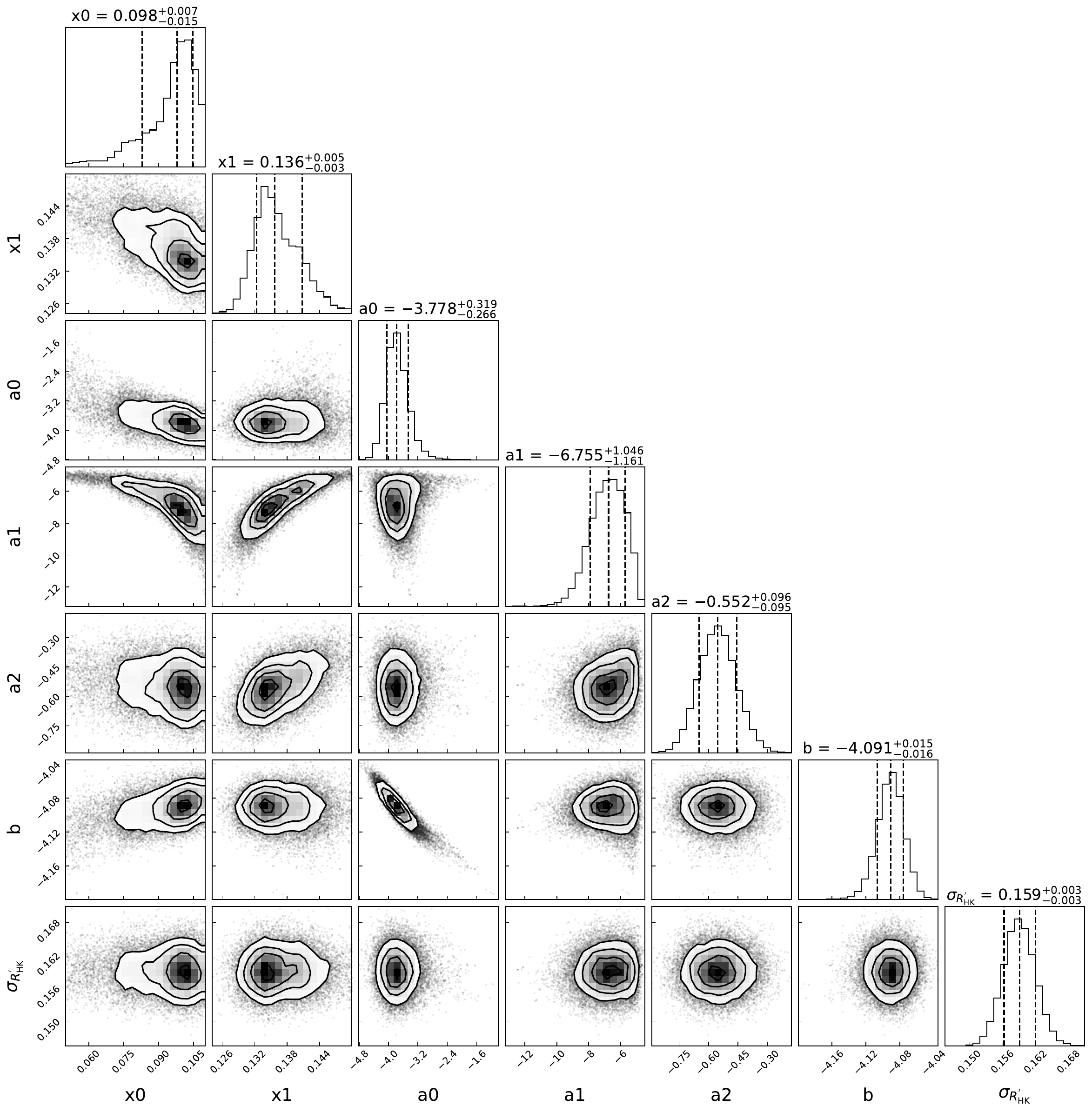}

\caption{Posterior probability distributions of the fitting results in linear-log scale corresponding to all points fitting method.}
\label{post_linear.fig}
\end{figure*}

\begin{figure*}[h]
\centering

\includegraphics[width=0.85\textwidth]{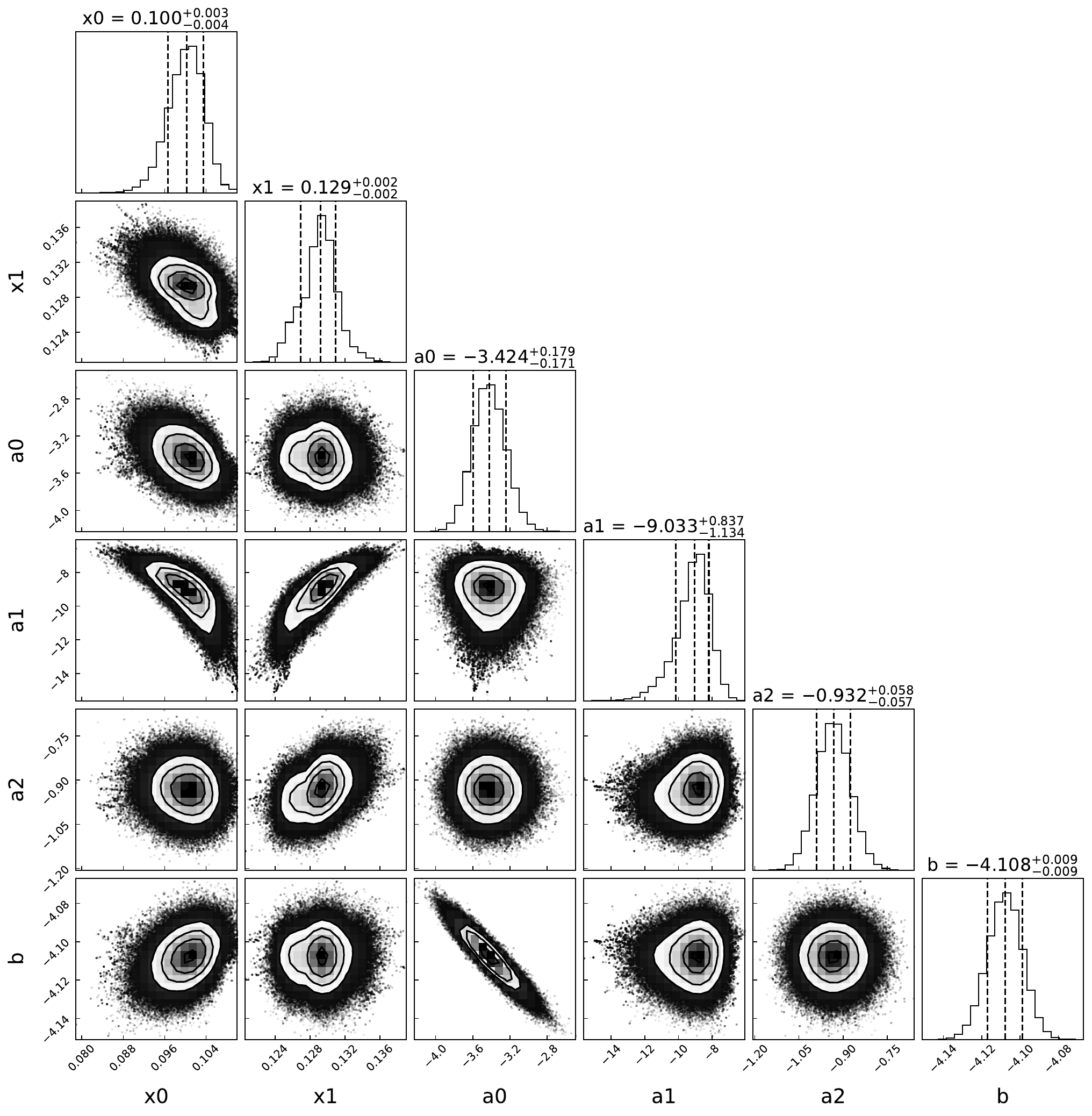}

\caption{Posterior probability distributions of the fitting results in linear-log scale corresponding to binned median fitting method.}
\label{post_linear_binned.fig}
\end{figure*}

We present the fitting results of rotation--activity relation in linear-log scale in Figure \ref{ar_fit_piece_linear.fig}. Because the first knee point can not be accurately resolved in linear scale, we fixed the point to be Ro $=$ 0.02 following previous studies \citep{2018A&A...618A..48M, 2025A&A...699A.251Y}. Posterior probability distributions of the fitting results are given in Figure \ref{post_linear.fig} and Figure \ref{post_linear_binned.fig}.

\clearpage

\section{Rotation--activity relations corresponding to different models and methods}
\renewcommand\thefigure{\Alph{section}\arabic{figure}}
\renewcommand\thetable{\Alph{section}\arabic{table}}
\setcounter{figure}{0}
\setcounter{table}{0}
In this section, we plot fitting results of rotation--activity relations corresponding to different models. Panel (a) shows the three-segment model while panel (b) shows the two-segment model. 

\begin{figure*}[h]
\centering
\subfigure[]{
\includegraphics[width=0.55\textwidth]{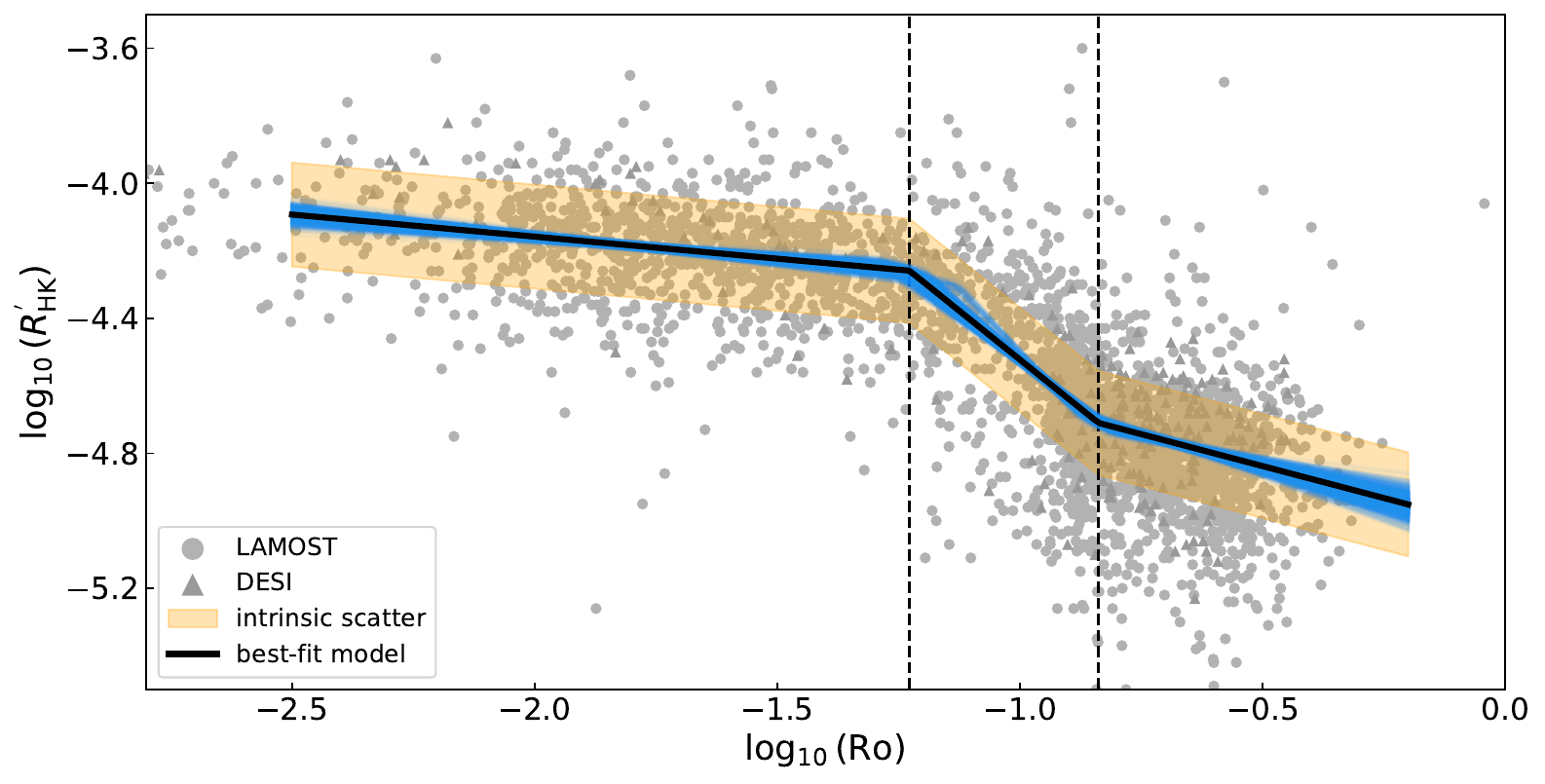}}
\subfigure[]{
\includegraphics[width=0.55\textwidth]{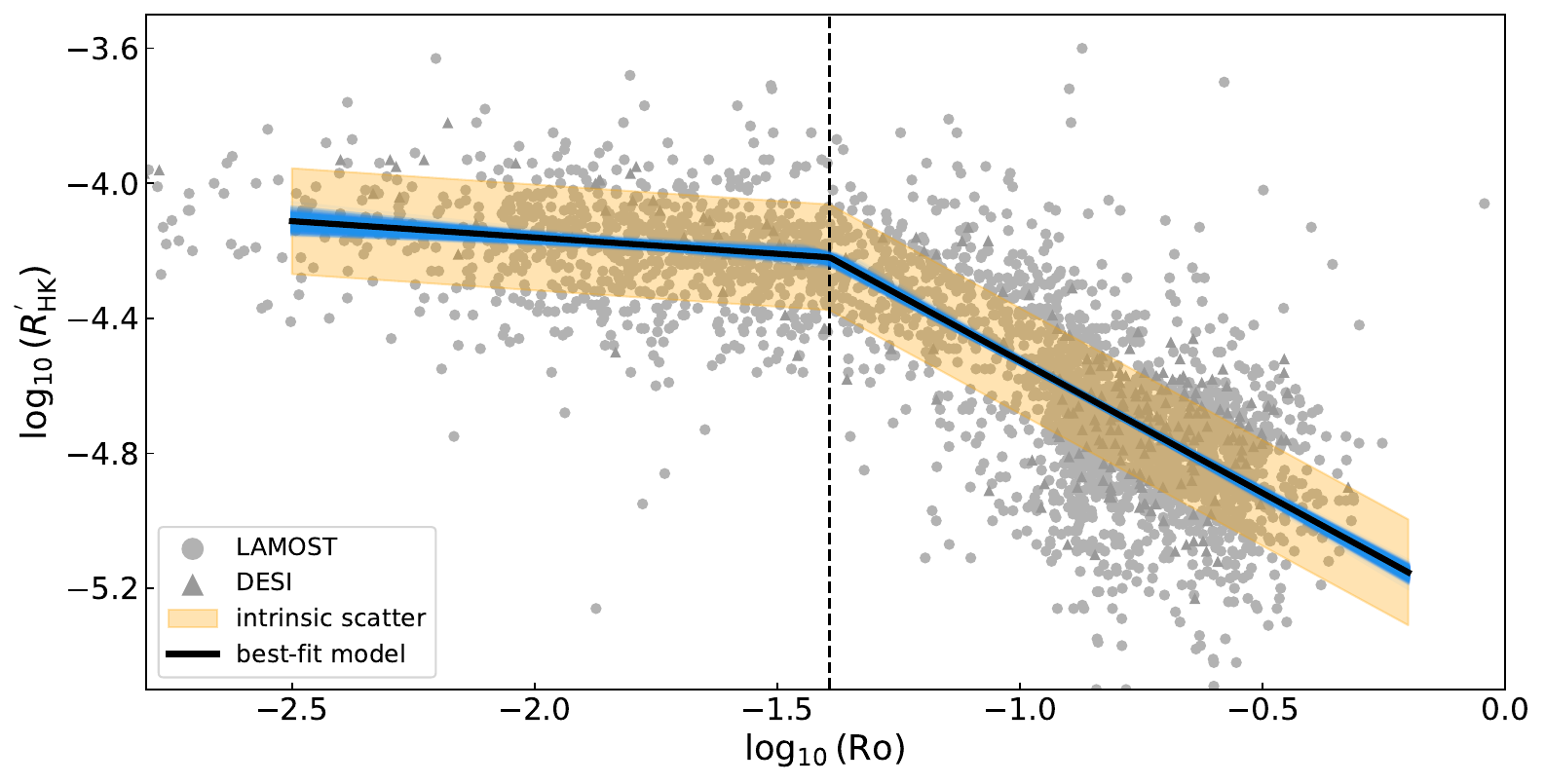}}
\caption{Panel (a): rotation--activity relation with three-segment model. Panel (b): rotation--activity relation with two-segment model.}
\label{ar_fit_various_models.fig}
\end{figure*}

\begin{figure*}[h]
\centering
\subfigure[]{
\includegraphics[width=0.55\textwidth]{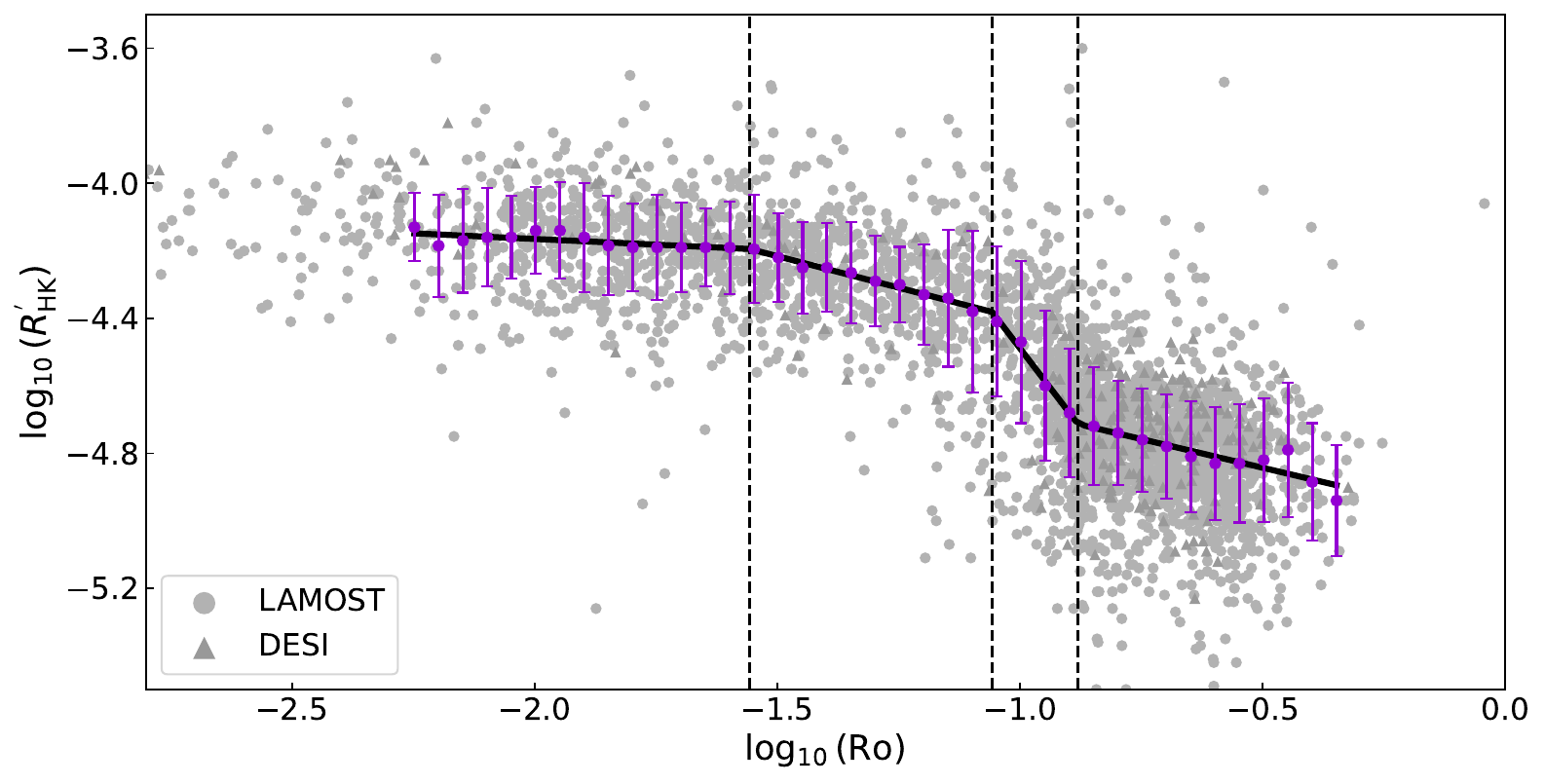}}
\caption{Four-segment fitting using the piecewise regression python package \citep{Pilgrim2021}.}
\label{ar_fit_various_P21.fig}
\end{figure*}

\clearpage

\end{appendix}



\end{document}